\documentclass[sigplan,nonacm]{acmart}

\acmConference[PLDI'20]{ACM SIGPLAN Conference on Programming Language Design and Implementation}{June 15--20, 2020}{London, UK}
\acmYear{2020}
\acmISBN{} %
\acmDOI{} %
\startPage{1}

\setcopyright{none}

\usepackage{booktabs}   %
\usepackage{subcaption} %
\usepackage{xspace}
\usepackage{microtype}
\usepackage{listings}

\begin{document}

\title[]{A Unified Iteration Space Transformation Framework\\for Sparse and Dense Tensor Algebra}

\author{Ryan Senanayake}
\affiliation{
  \institution{MIT CSAIL, USA}            %
}
\email{rsen@mit.edu}          %

\author{Fredrik Kjolstad}
\affiliation{
  \institution{MIT CSAIL, USA}
}
\email{fred@csail.mit.edu}

\author{Changwan Hong}
\affiliation{
  \institution{MIT CSAIL, USA}
}
\email{changwan@mit.edu}

\author{Shoaib Kamil}
\affiliation{
  \institution{Adobe, USA}
}
\email{kamil@adobe.com}

\author{Saman Amarasinghe}
\affiliation{
  \institution{MIT CSAIL, USA}
}
\email{saman@csail.mit.edu}

\definecolor{todocolor}{rgb}{0.8,0,0}

\newcommand{\TODO}[1]{{\color{todocolor}#1}}
\newcommand{\figref}[1]{Figure~\ref{fig:#1}}
\newcommand{\figsref}[2]{Figures~\ref{fig:#1} and~\ref{fig:#2}}
\newcommand{\subfigref}[1]{(\subref{fig:#1})}
\newcommand{\secref}[1]{Section~\ref{sec:#1}}
\newcommand{\subsecref}[1]{\S\,\ref{sec:#1}}
\newcommand{\eqnref}[1]{Eq.~\ref{eqn:#1}}
\newcommand{\eqnsref}[2]{Eqs.~\ref{eqn:#1} and~\ref{eqn:#2}}
\newcommand{\eqnrref}[2]{Eqs.~\ref{eqn:#1}--\ref{eqn:#2}}
\newcommand{\defref}[1]{Definition~\ref{def:#1}}
\newcommand{\tabref}[1]{Table~\ref{tab:#1}}
\newcommand{\HIDE}[1]{}

\definecolor{keywordcolor}{rgb}{0.5,0,0.5}
\definecolor{textgray}{gray}{0.4}
\definecolor{mygray}{rgb}{0.5,0.5,0.5}
\lstset {
  language=C++,
  columns=fullflexible,
  numbers=none,
  numbersep=5pt,
  numberstyle=\scriptsize\color{mygray},
  keywordstyle=\color{keywordcolor},
  escapeinside={(*}{*)}
}
\newcommand\code[1]{\lstinline[mathescape=true]|#1|}

\newcommand{\spmv}{\texttt{SpMV}\xspace}

\begin{abstract}

  We address the problem of optimizing mixed sparse and dense
  tensor algebra in a compiler.  We show that standard loop
  transformations, such as strip-mining, tiling, collapsing,
  parallelization and vectorization, can be applied to irregular
  loops over sparse iteration spaces.  We also show how these
  transformations can be applied to the contiguous value arrays of
  sparse tensor data structures, which we call their position
  space, to unlock load-balanced tiling and parallelism.

  We have prototyped these concepts in the open-source TACO
  system, where they are exposed as a scheduling API similar to
  the Halide domain-specific language for dense computations.
  Using this scheduling API, we show how to optimize mixed
  sparse/dense tensor algebra expressions, how to generate
  load-balanced code by scheduling sparse tensor algebra in
  position space, and how to generate sparse tensor algebra GPU
  code.  Our evaluation shows that our transformations let us
  generate good code that is competitive with many hand-optimized
  implementations from the literature.

\end{abstract}

\keywords{Sparse Tensor Algebra, Optimization}  %

\maketitle

\section{Introduction}
\label{sec:introduction}

There has been a lot of interest in compilers for
dense~\cite{lgen, btoblas, auer2006, ragan-kelley2012,
vasilache2018} and
sparse~\cite{kjolstad2017,chou2018,kjolstad2019} linear and tensor
algebra.  Dense tensor algebra compilers, such as TCE, TVM and
Tensor Comprehensions, build on decades of research on loop
transformations for affine loop nests~\cite{wolfe1982,
wolf1991data, mckinley1996improving}.
Sparse tensor algebra compilers, on the other hand, suffer from a
lack of analogous sparse loop transformation frameworks.

Without a loop transformation framework, sparse tensor algebra
compilers leave several optimization opportunities on the table.
First, sparse tensor algebra expressions are often really a mix of
dense and sparse tensor algebra, where some operands are stored in
sparse data structures and some in dense arrays.  Examples include
SpMV with a dense vector and (sparse matrix) $\times$
(dense matrix) multiplication (SpMM).  Second, sparse
tensors often have some dimensions that are dense, such as the
blocked compressed sparse rows matrix format (BCSR) representing a
sparse matrix that has dense blocks.  Without a loop
transformation framework, sparse tensor algebra compilers cannot
optimize the dense loops in mixed sparse and dense expressions.
Furthermore, current sparse tensor algebra compilers cannot apply
loop transformations, such as strip-mine, reorder and collapse, to
sparse loops.  We will show that such transformations are possible
and that they can be used to enable parallelization and make
effective use of modern GPUs.  Finally, we will show that by
tiling the contiguous nonzero value arrays in sparse tensors, we
can generate static load-balanced parallel code.

\begin{figure}
  \includegraphics[width=0.90\linewidth]{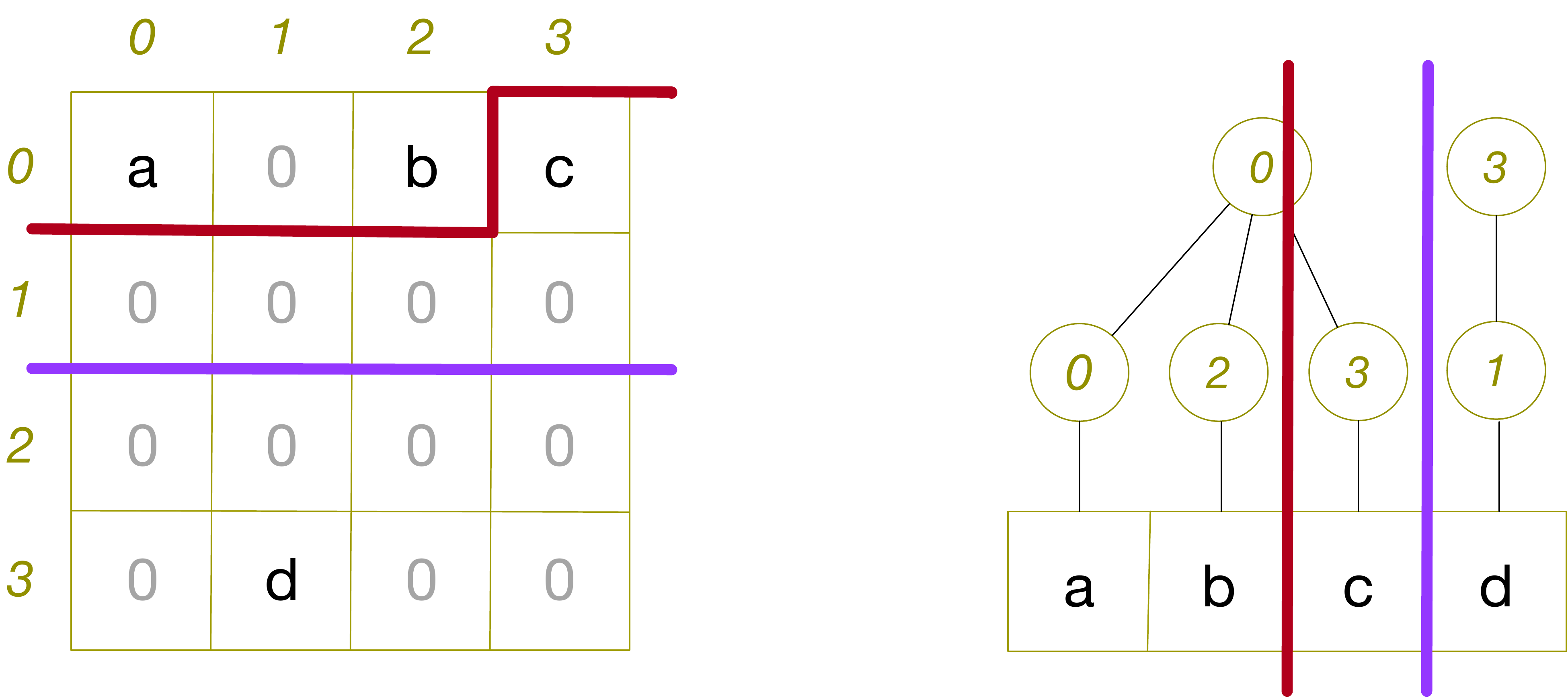}
  \caption {
    \label{fig:cuts}
    A coordinate cut (purple) and a position cut (red) in the
    coordinate iteration space of a matrix (left), and the same
    cuts in its compressed coordinate data structure.  The
    coordinate cut cleanly divides the iteration space, while the
    position cut cleanly divides the values of the data structure.
  }
\end{figure}

Applying loop transformations to sparse loops is challenging
because they iterate over data structures.  These data structures
encode the coordinates of nonzero values and a single sparse loop
may iterate over any number of them.  In fact, the resulting
imperative code may contain while loops, conditionals and indirect
memory references.  Furthermore, tiling loop transformations rely
on the ability to jump around the iteration space, and sparse data
structures may not support random access.

In this paper, we propose a unified loop transformation framework
for loop nests with both dense and sparse loops that come from
sparse tensor algebra.  The loop transformations are applied to
intermediate representations before generating imperative code.
The transformations let us strip-mine, reorder, collapse,
vectorize, and parallelize both dense and sparse loops subject to
straightforward preconditions.  Furthermore, our transformations enable 
sparse loops to be strip-mined both in the normal iteration space
corresponding to dense loop nests and in the contiguous space
given by the nonzeros of its compressed data structure
(\figref{cuts}).

\newcommand{\codescale}{0.48}
\begin{figure*}
  \begin{minipage}{0.36\linewidth}
    \begin{minipage}{\linewidth}
      \includegraphics[scale=\codescale]{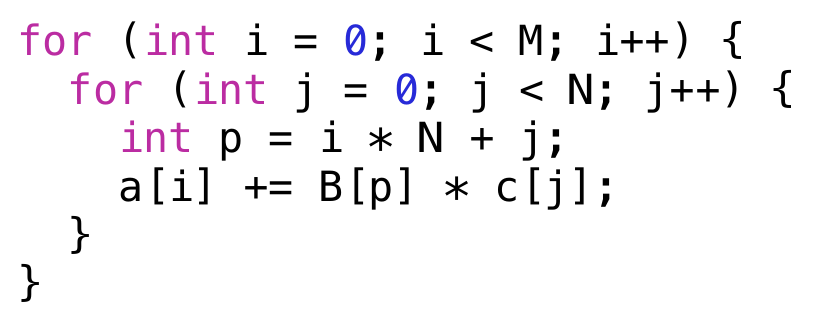}
      \subcaption {
        Unscheduled GEMV
        \label{fig:example-code-dense}
        \vspace{7mm}
      }
    \end{minipage}
    \begin{minipage}{\linewidth}
      \includegraphics[scale=\codescale]{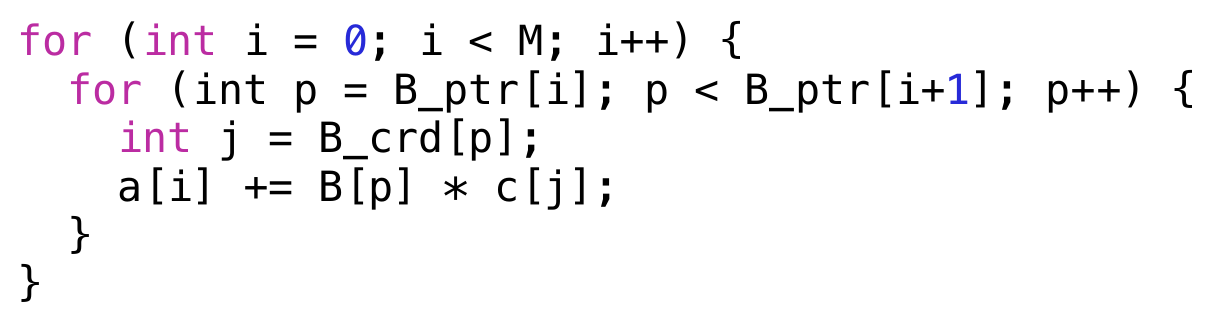}
      \subcaption {
        Unscheduled SpMV
        \label{fig:example-code-unscheduled}
        \vspace{7mm}
      }
    \end{minipage}
    \begin{minipage}{\linewidth}
      \includegraphics[scale=\codescale]{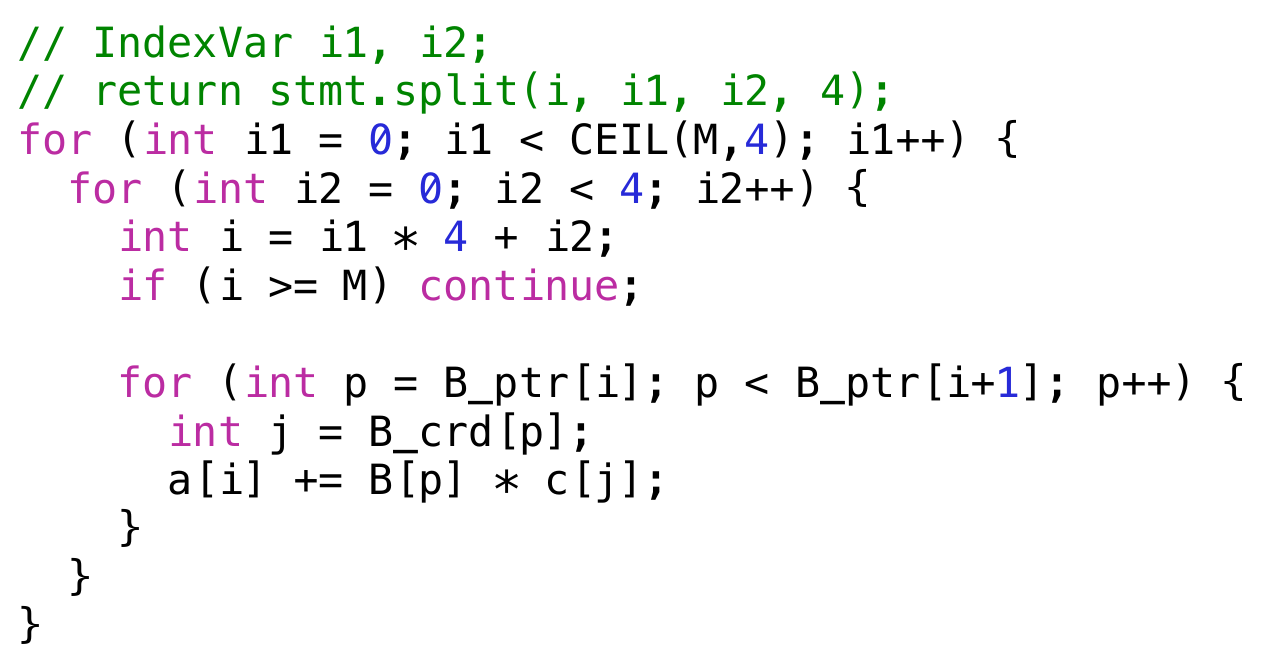}
      \subcaption {
        Strip-mined SpMV
        \label{fig:example-code-tiled}
        \vspace{6mm}
      }
    \end{minipage}
    \begin{minipage}{\linewidth}
      \includegraphics[scale=\codescale]{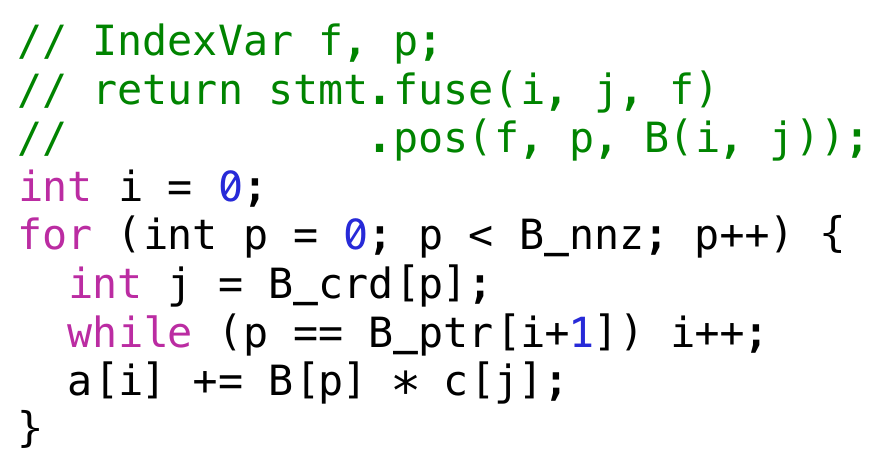}
      \subcaption {
        Position Iterating SpMV
        \label{fig:example-code-pos}
      }
    \end{minipage}
  \end{minipage}
  \begin{minipage}{0.63\linewidth}
    \includegraphics[scale=\codescale]{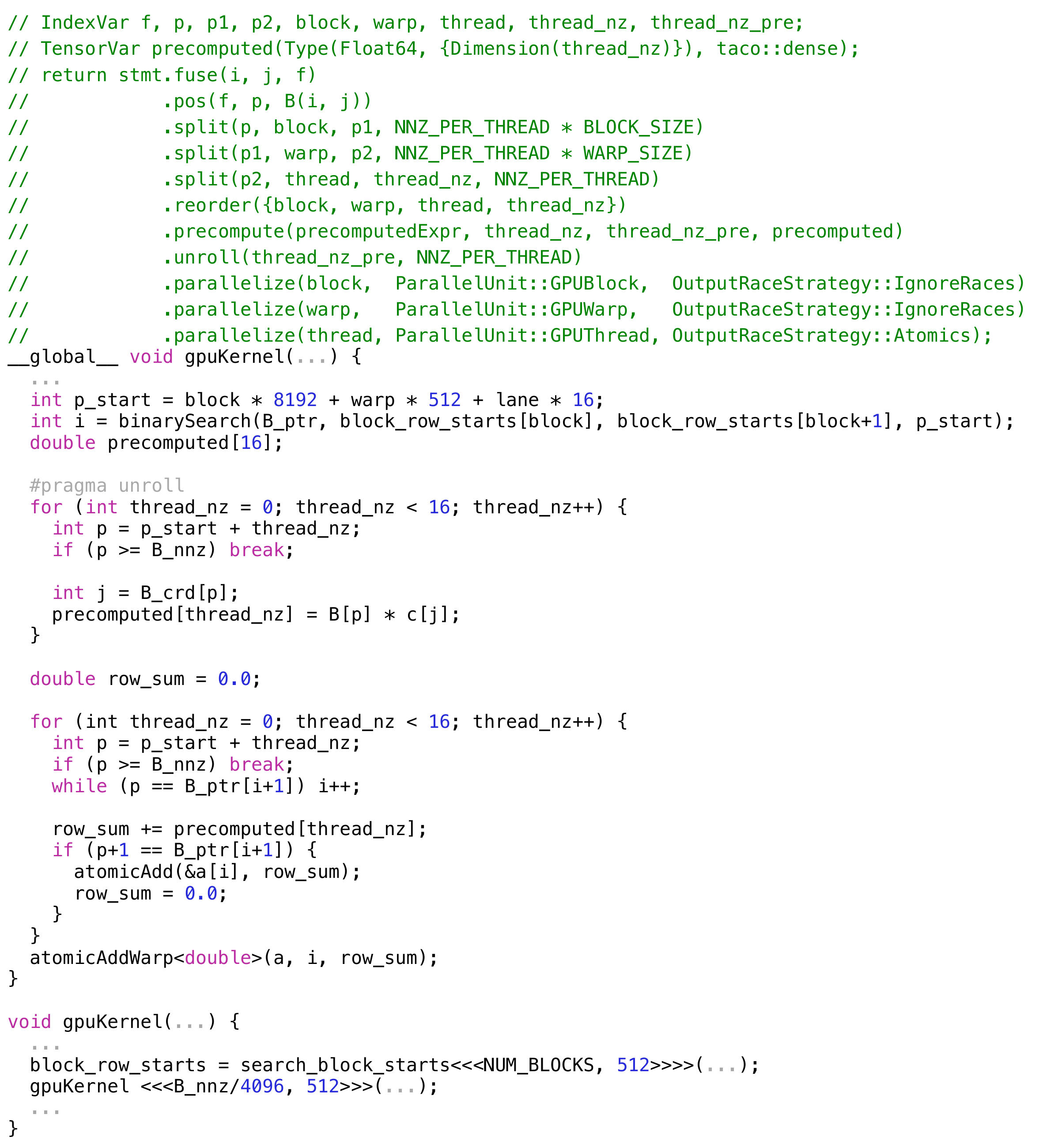}
    \subcaption {
      GPU Optimized SpMV
      \label{fig:example-code-gpu}
    }
  \end{minipage}
  \caption{
    Matrix-vector multiplication computed in several different
    ways, controlled by sparse iteration space transformations.
    \label{fig:spmv-examples}
  }
\end{figure*}

We have prototyped these ideas in the open-source TACO sparse
tensor algebra compiler of Kjolstad et al.~\cite{kjolstad2017},
which can generate code for any sparse tensor algebra expression.
We expose the transformations as a loop scheduling API analogous
to the Halide scheduling language~\cite{ragan-kelley2012}.  Unlike
Halide, however, the transformations can apply to both sparse and
dense loops from mixed sparse and dense tensor algebra
expressions.  We leave automatic scheduling---developing an
optimization system to automatically decide loop
transformations---as future work.  Such systems can be built on
top of the transformations we propose here.  Finally, we show that
our transformations let us generate good code that is competitive
with hand-optimized kernels where these are available, that they
let us target GPUs, and that they let us generate load-balanced
kernels.

Our specific contributions are
\begin{enumerate}

\item position iteration spaces to complement the standard
coordinate iteration spaces,

\item generalization of standard loop transformations to sparse
loops,

\item dimension relationship graphs to track the dependencies 
between iteration spaces and data structures, and

\item a CUDA GPU backend for sparse tensor algebra enabled by
sparse loop transformations.

\end{enumerate}

\section{SpMV Example}
\label{sec:motivating-example}

In this section, we demonstrate the capabilities of our loop
transformation and code generation framework using a motivating
example.  Our main target is code expressing computations on mixed
dense and sparse multidimensional tensors, but for simplicity we
use the well-known example of matrix-vector multiplication:
$a=Bc$, in which a vector $c$ (a one-dimensional tensor) is
multiplied by a matrix $B$ (a two-dimensional tensor).

When all operands are dense, the code is straightforward.
Figure~\ref{fig:spmv-examples}a shows matrix-vector multiplication
with dense operands; the loops shown are easy to optimize, since
they iterate over a regular data structure.  However, when the
matrix is sparse, as shown in Figure~\ref{fig:spmv-examples}b,
iterating through the entries of the matrix becomes more complex.
This is due to the fact that sparse matrices are most often stored
in \textit{compressed} structures\footnote{In this example, we use
the ubiquitous compressed sparse rows storage.}, which avoid
storing zero entries (more details about matrix storage are given
in Section~\ref{sec:positionspace}).  The combination of complex
iteration and indirect storage, common to most sparse matrix
formats, makes subsequent loop transformations more difficult for
sparse matrix-vector multiplication (\spmv).

In the simplest case, shown in Figure~\ref{fig:spmv-examples}c, it
is possible to tile \spmv by tiling the outer dense loop.  This
code is suitable for further optimizations such as parallelizing
the loop over \texttt{i2}.  However, the resulting code can be
load-imbalanced, because different rows may contain different
numbers of entries.  Logically, this approach tiles the $i$ outer
dimension of the matrix $B$ while iterating through the $i$ and
$j$ dimensions of the matrix.  We refer to this approach as
iterating through the \textit{coordinate space} of $B$.

The alternative approach is to instead iterate through the
nonzeros of $B$, which we call iterating through the
\textit{position space}.  Code for computing \spmv by iterating
through position space is shown in
Figure~\ref{fig:spmv-examples}d.  This style of iteration is
statically load-balanced and can be used as the basis for high
performance implementations on GPUs.
Figure~\ref{fig:spmv-examples}e shows an optimized GPU
implementation of \spmv, which iterates through position space.
This code requires a number of further transformations, including
further tiling for GPU blocks, warps, and threads, as well as loop
unrolling and precomputation.  Transforming \spmv to iterate
through position space, however, is the basis on which the rest of
the transformations are applied.

While the code in Figure~\ref{fig:spmv-examples}a-b can be
generated by TACO, the extensions in this work enable the compiler
to automatically generate the optimized implementations shown in
the rest of the figure.  In the rest of the paper, we describe how
we automate transforming code to iterate through position space
for sparse tensors, and present a scheduling language users can
use to explicitly apply this transformation along with others
required to generate high performance sparse tensor code on CPUs
and GPUs.

\section{Coordinate \& Position Iteration Spaces}
\label{sec:positionspace}
There are two types of iteration spaces: coordinate spaces and
position spaces.  A coordinate space is a multi-dimensional
Cartesian combination of coordinates that encode each dimension.
Position spaces, on the other hand, are the positions along the
space-filling curve created by imposing an ordering on these
coordinates.  We can imagine many different orderings, but in this
paper we will limit ourselves to lexicographical orderings.

\begin{figure}
  \begin{minipage}{\linewidth}
    \center
    \begin{minipage}[b]{0.35\linewidth}
      \center
      \includegraphics[width=.9\linewidth]{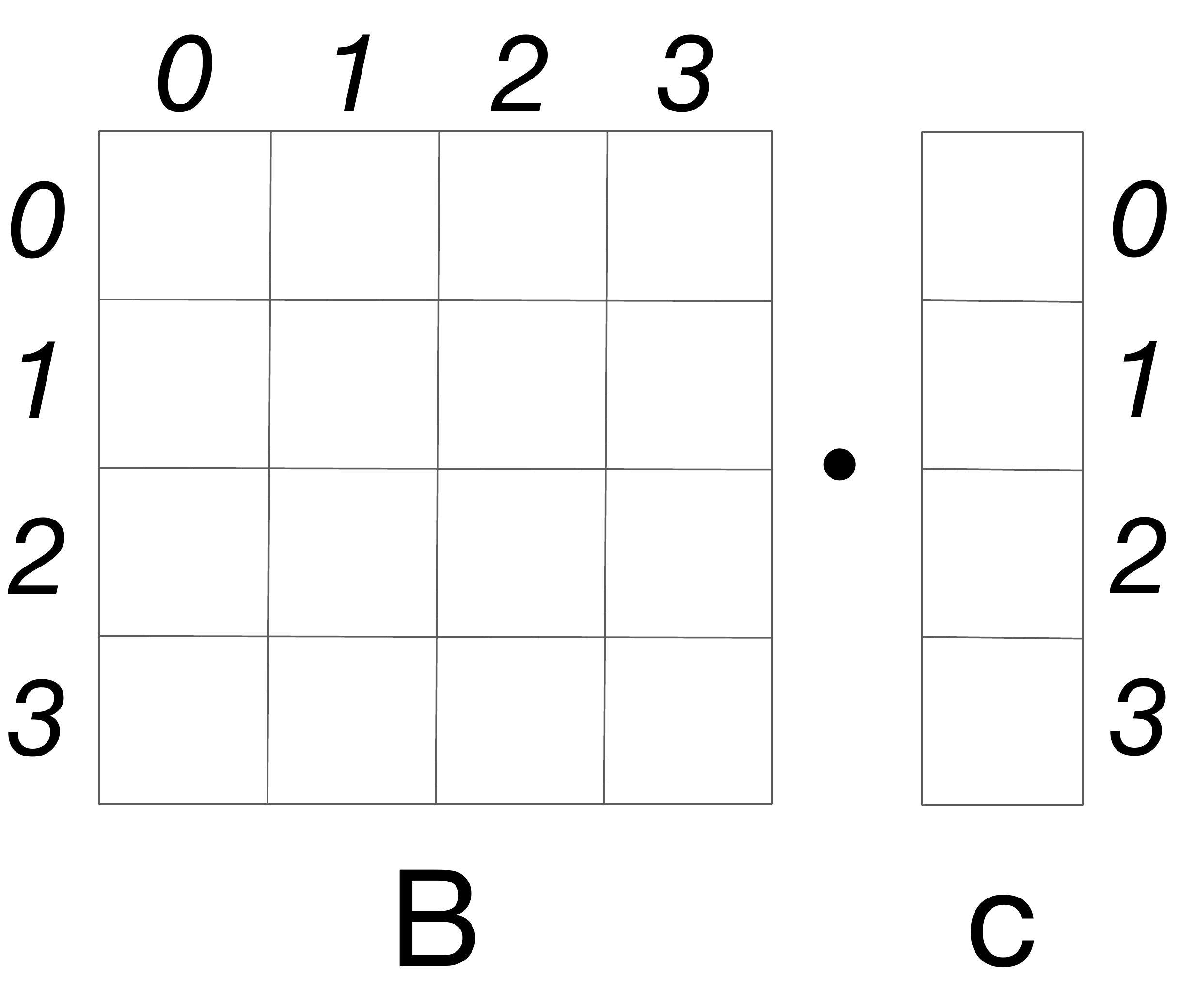}
      \subcaption{
        Data structures
        \label{fig:space-dense-datastructure}
      }
    \end{minipage}
    \begin{minipage}[b]{0.35\linewidth}
      \center
      \includegraphics[width=.6\linewidth]{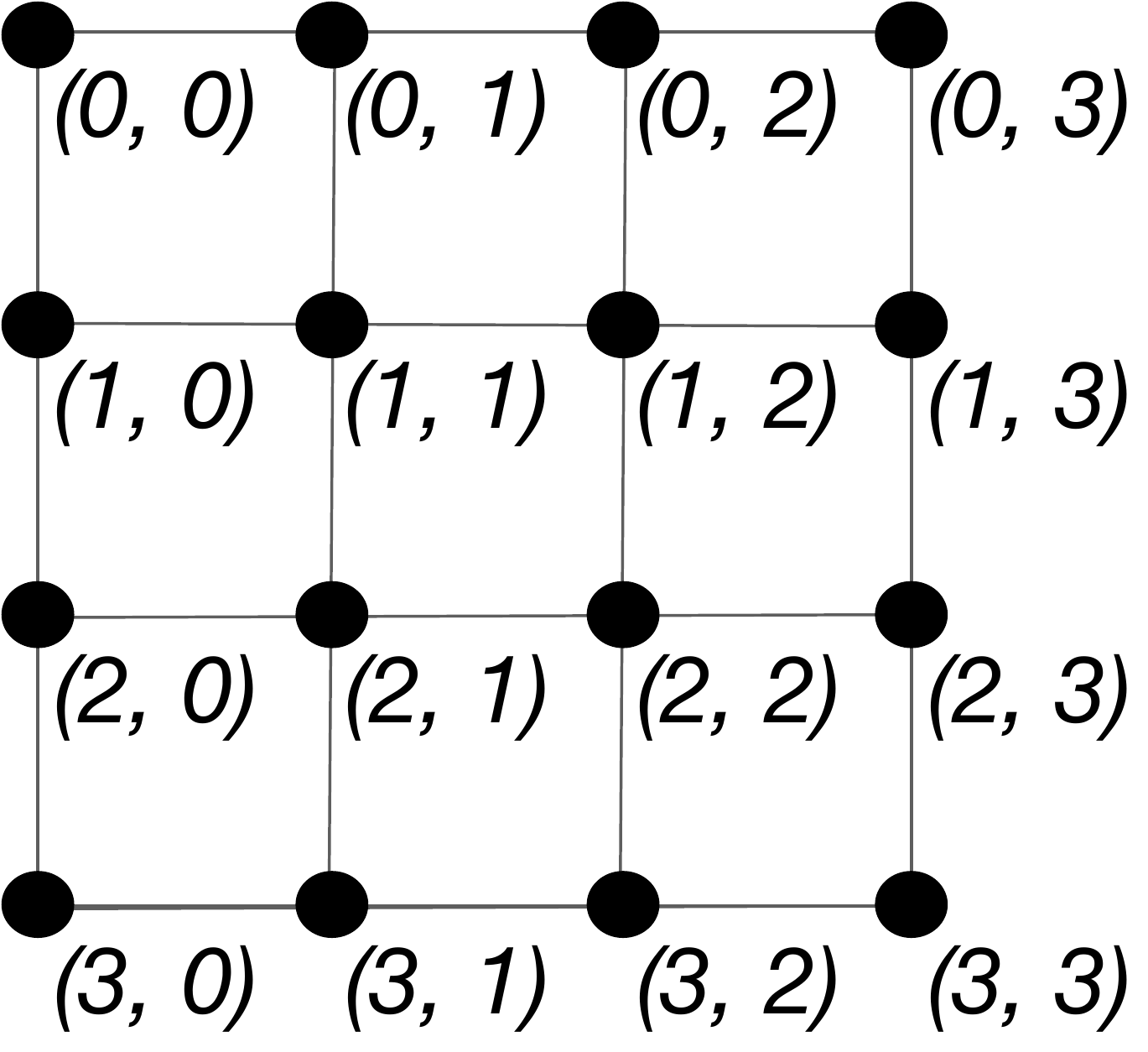}
      \subcaption{
        Coordinates
        \label{fig:space-dense-coordinates}
      }
    \end{minipage}
    \begin{minipage}[b]{0.25\linewidth}
      \center
      \includegraphics[width=\linewidth]{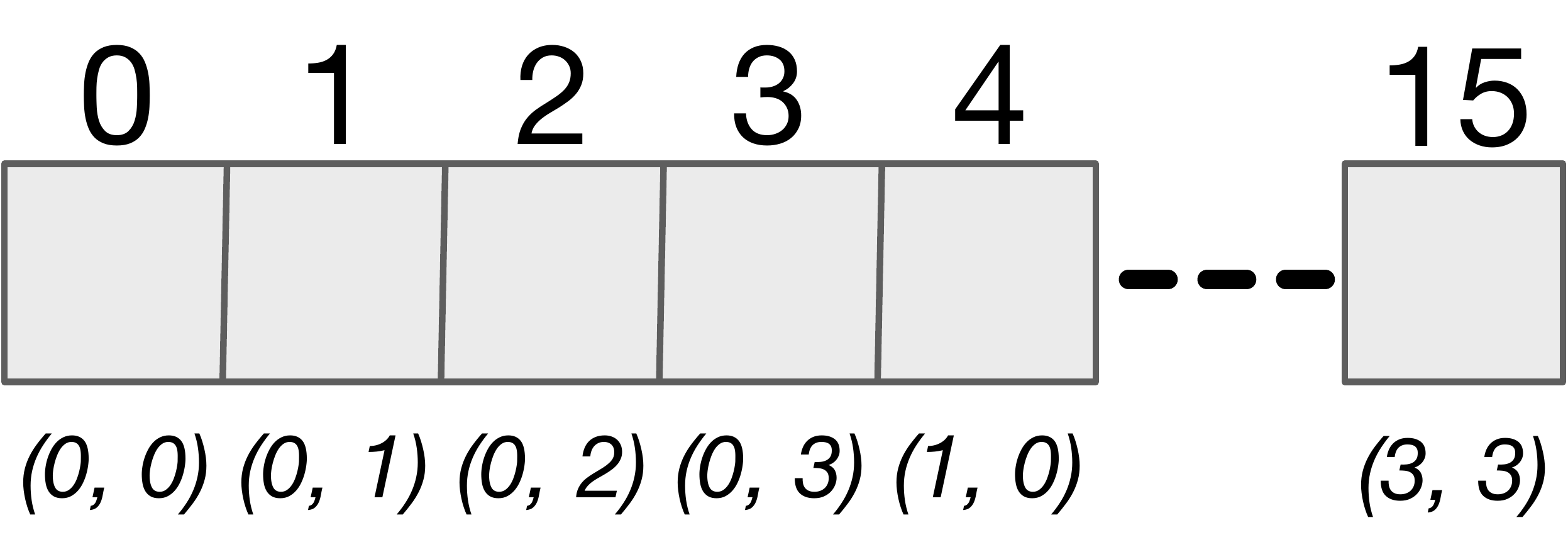}
      \vspace{0.01mm}
      \subcaption{
        Positions
        \label{fig:space-dense-positions}
      }
    \end{minipage}
    \caption {
      The dense matrix-vector multiplication ($a=Bc$) data
      structures, coordinate space and position space.
      \vspace{4mm}
    }
  \end{minipage}
  \begin{minipage}{\linewidth}
    \center
    \begin{minipage}[b]{0.35\linewidth}
      \center
      \includegraphics[width=0.9\linewidth]{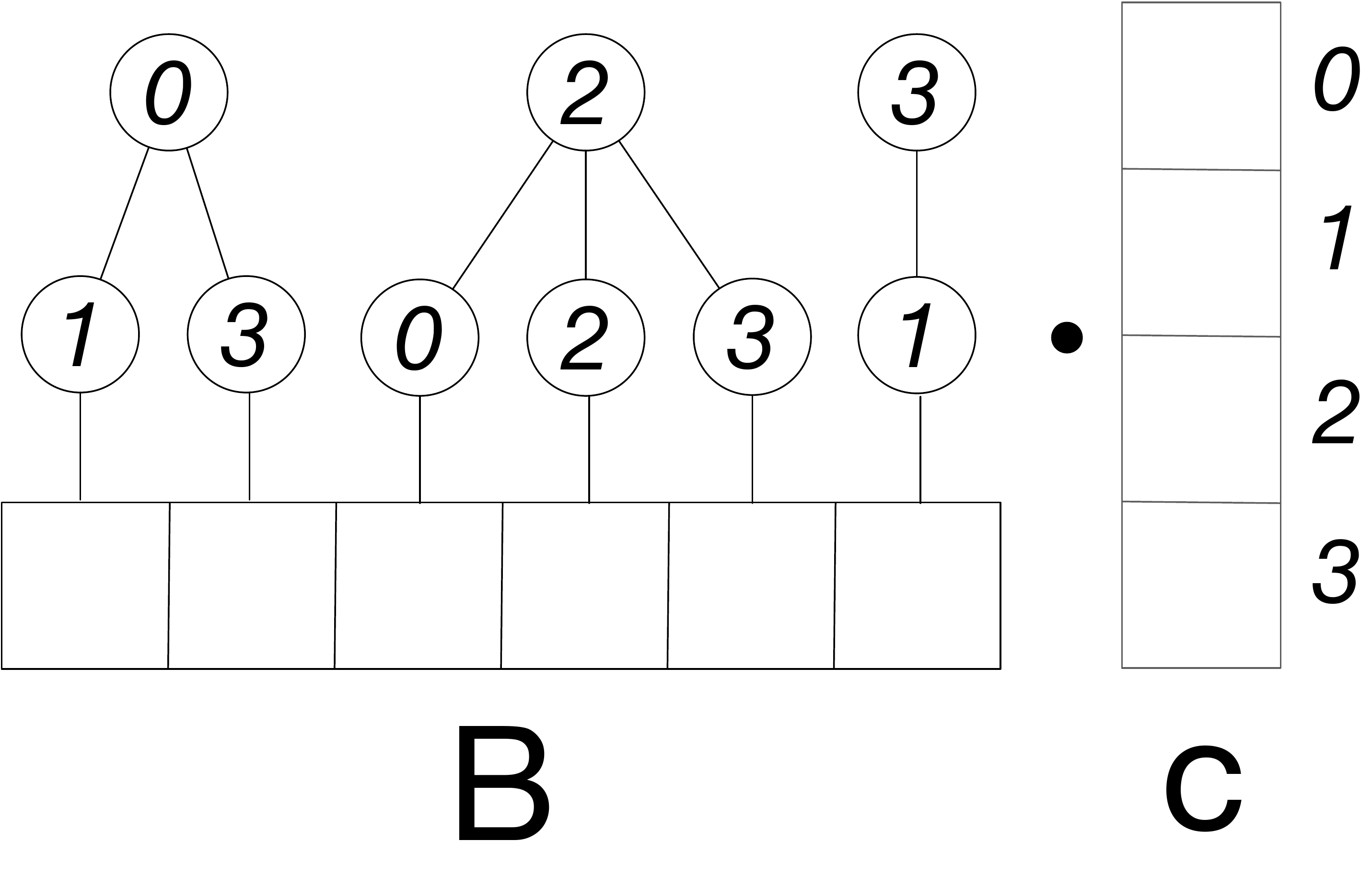}
      \subcaption{
        Data structures
        \label{fig:space-sparse-datastructure}
      }
    \end{minipage}
    \begin{minipage}[b]{0.35\linewidth}
      \center
      \includegraphics[width=0.6\linewidth]{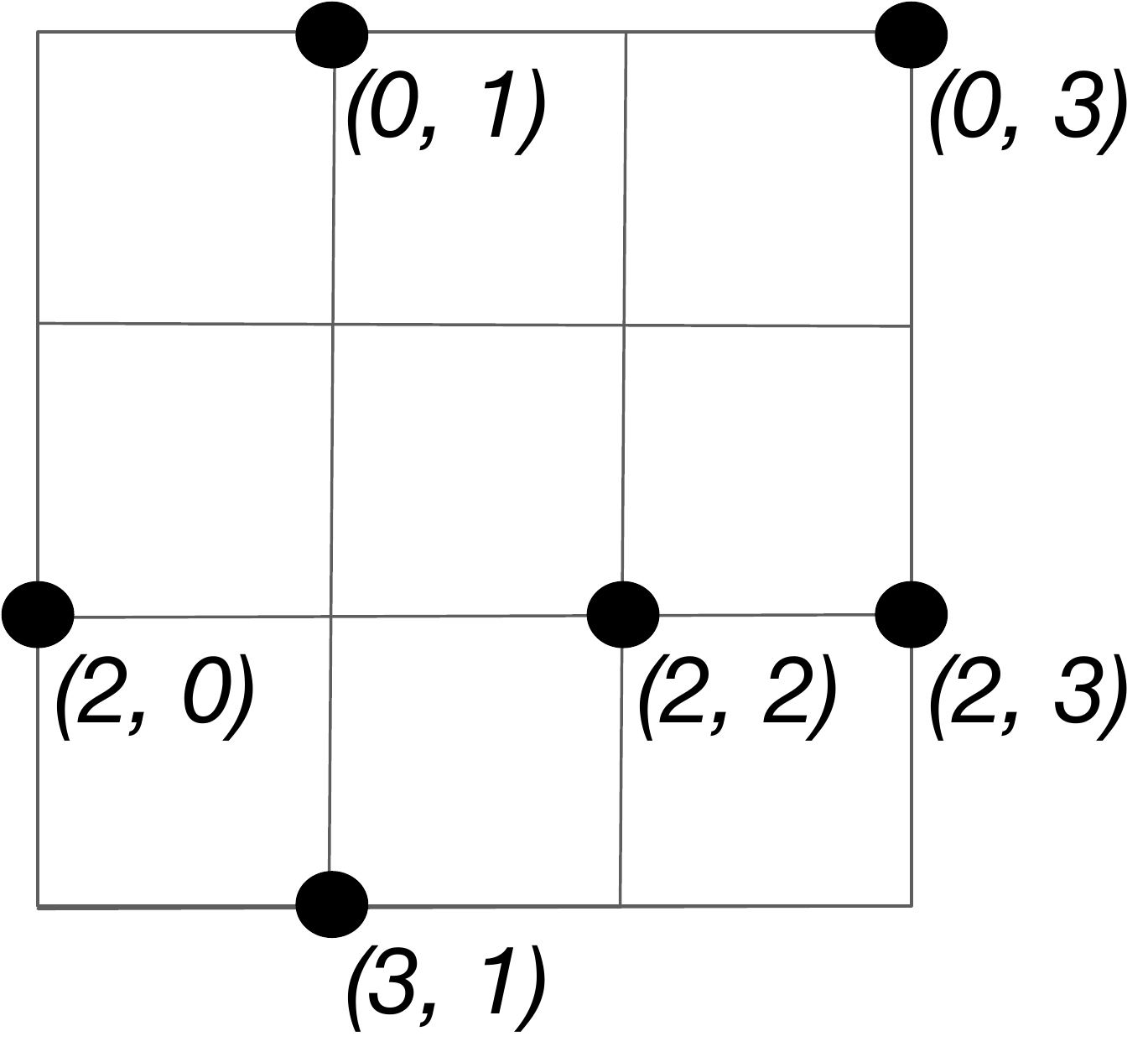}
      \subcaption{
        Coordinates
        \label{fig:space-sparse-coordinates}
      }
    \end{minipage}
    \begin{minipage}[b]{0.25\linewidth}
      \center
      \includegraphics[width=0.9\linewidth]{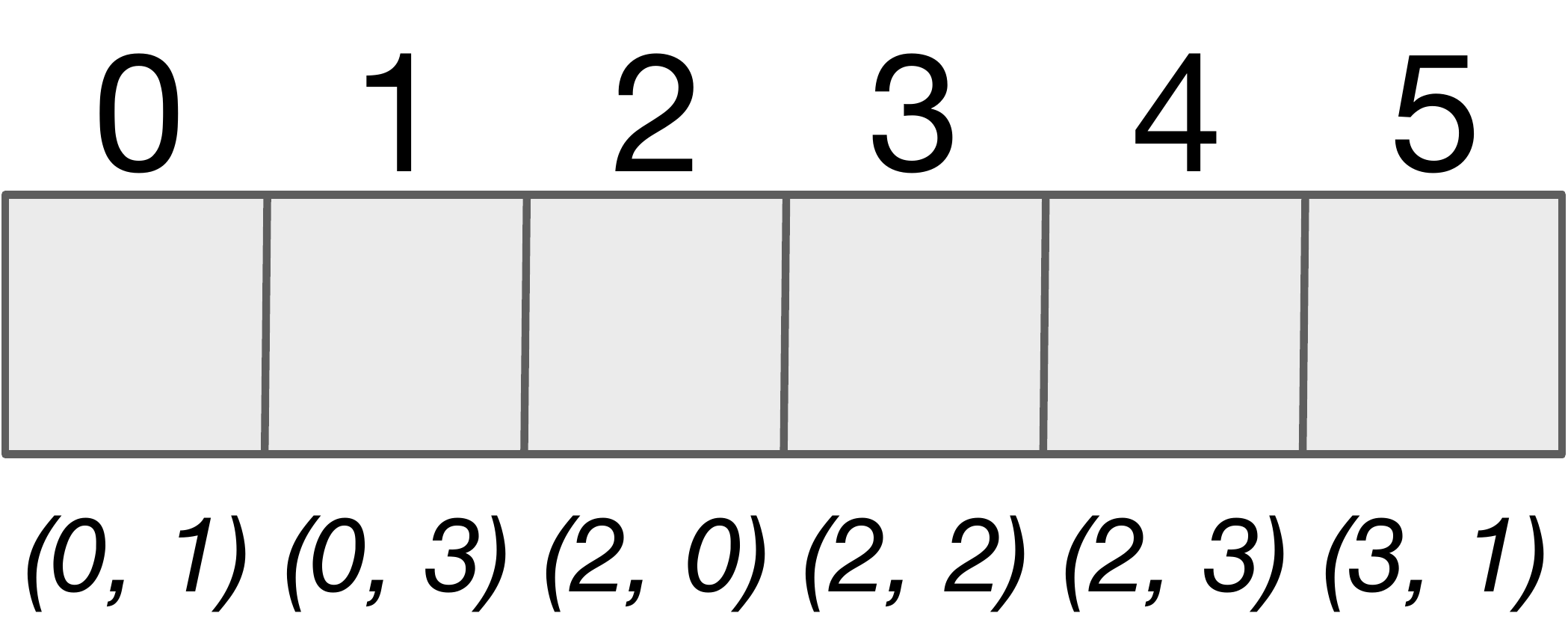}
      \vspace{0.01mm}
      \subcaption{
        Positions
      \label{fig:space-sparse-positions}
    }
    \end{minipage}
    \caption {
      The sparse matrix-vector multiplication (\spmv) data
      structures, coordinate space and position space.
    }
  \end{minipage}
\end{figure}

Dense coordinate spaces, such as those that arise from dense
linear and tensor algebra, can be visualized as a
multi-dimensional lattice (i.e., a grid).
\figref{space-dense-datastructure} shows a dense matrix-vector
multiplication example $a=Bc$, and
\figref{space-dense-coordinates} its two-dimensional $m \times n$
coordinate space, lexicographically ordered with the $i$
coordinates before the $j$ coordinates.  The resulting loop nest,
shown in~\figref{example-code-dense}, iterates over $(i,j)$
coordinates and computes a position \code{p} to access the matrix
using a strided formula $i*n + j$.  \figref{space-dense-positions}
shows the multiplication's one-dimensional position space,
consisting of the positions along the lexicographical ordering of
the coordinate space.  The resulting loop nest would iterate over
these positions $p$ and compute the coordinates with the formulas
$i = p/N$ and $j = p\%N$.  These would, of course, be expensive to
compute and position iteration over a dense matrix therefore
primarily makes sense when the coordinates are not needed, such as
when scaling a matrix.

Sparse iteration spaces appear when a loop iterates over one or
more data structures that encode subsets of the coordinates in an
iteration space dimension.  In sparse tensor algebra, one or more
tensors are stored in hierarchical compressed data structures.
\figref{space-sparse-datastructure} shows an example of a matrix
$B$ whose coordinates are stored in a compressed hierarchy 
that contains only those coordinates that lead to a nonzero matrix
value.  Note that, although we chose to show the coordinate
hierarchy abstractly as a forest, in computer memory it would be
stored as a concrete data structure, such as compressed sparse
rows (CSR) or doubly-compressed sparse rows (DCSR).  See
Kjolstad et al.~\cite{kjolstad2017} or Chou et
al.~\cite{chou2018} for in-depth descriptions of the relationships
between coordinate hierarchies and concrete matrix/tensor data
structures.

Sparse coordinate spaces can be visualized as a multi-dimensional
lattice with holes.  The holes appear because the coordinate
hierarchy data structures compress out coordinates, and the sparse
iteration space is described by the coordinates that are visited
when loops iterate over these data structures.
\figref{space-sparse-coordinates} shows the sparse iteration space
of the \spmv operation.  The resulting loop nest that iterates over
it is shown in~\figref{example-code-unscheduled}.  The outer loop
iterates over the nonempty rows of \code{B}, stored as
$(\text{coordinate},\text{position})$ pairs.  The positions are
the locations of the coordinate at that level of the hierarchy, so
the \code{i} coordinate $2$ is stored at the second position after
coordinate $0$.  The inner loop iterates over the nonempty
(nonzero) components of the current row in the outer loop.  The
loops together iterates over the coordinate hierarchy of \code{B}
and thus the sparse iteration space of $Bc$.  (It is not necessary
to iterate over \code{c} as the intersection resulting from the
multiplication makes it is sufficient to iterate over the smaller
operand.)

Sparse position spaces are the sequence of nonzero coordinates in
the order they are stored in a coordinate hierarchy, which may be
thought of as a sparse space-filling curve through nonzero values.
Their main advantage is that, although the coordinate space of a
sparse tensor expression is sparse and irregular, its position
space is a dense one-dimensional space that can be effectively
tiled into equal-size blocks.  As we shall see, this makes it
possible to transform tensor expressions into statically
load-balanced parallel code and to make effective use of
vectorization and GPUs.  \figref{space-sparse-positions} shows the
position space of the \spmv expression as a sequence of positions
with coordinates attached.

\figref{example-code-pos} shows the single loop that iterates
through the sparse position space of the \spmv operation.  Like the
dense position space, the loop increments a position variable
\code{p}.  Based on the position it retrieves the coordinate
\code{j} from the matrix \code{B}'s coordinate array.  In
addition, it keeps track of the current coordinate \code{i} by
incrementing it every time it reaches the end of a row.  The
increment of \code{i} is placed in a while loop to increment past
empty rows.  This increment strategy requires that the coordinates
are stored in order.  If they are not, the increment of \code{i}
would be replaced by a search.  If there were more dimensions to
the matrix \code{B} then the code would keep track of each
coordinate above \code{i} using the same strategy.

\begin{figure}
  \begin{minipage}{\linewidth}
    \centering
    \includegraphics[width=\linewidth]{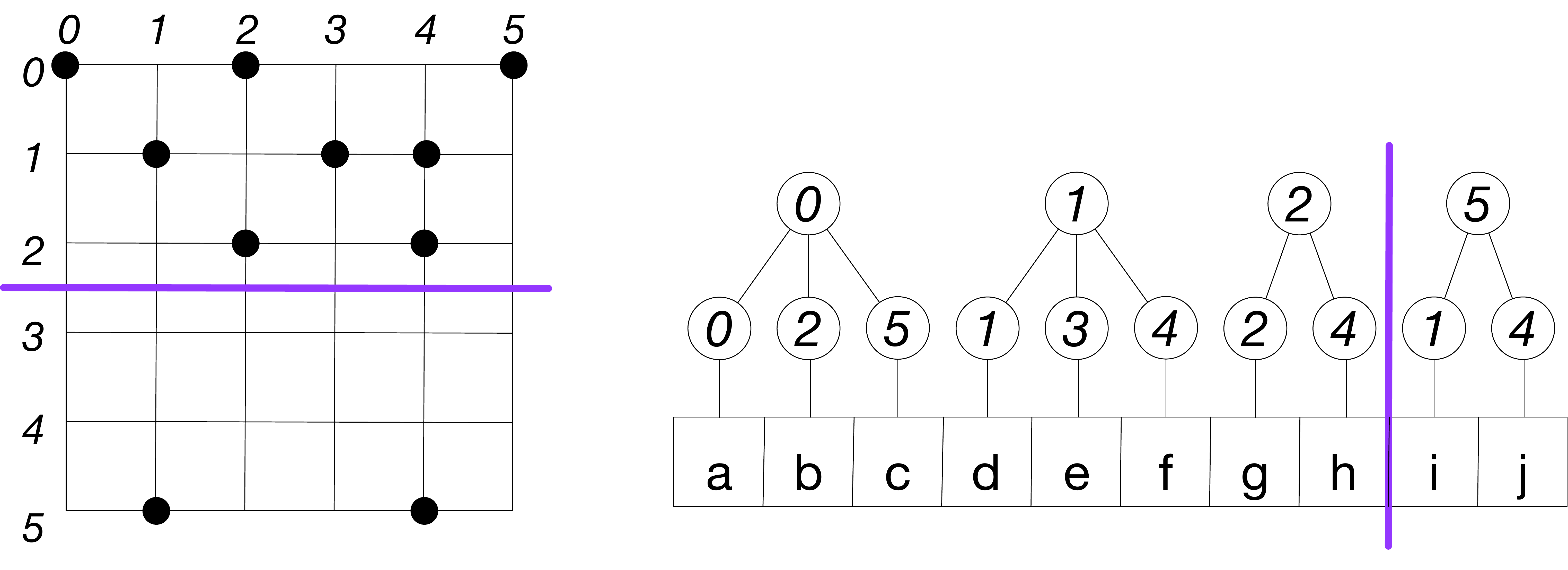}
    \caption {
      Tiling the coordinate space.
      \label{fig:coordinate-cut}
    }
  \end{minipage}
  \begin{minipage}{\linewidth}
    \centering
    \includegraphics[width=\linewidth]{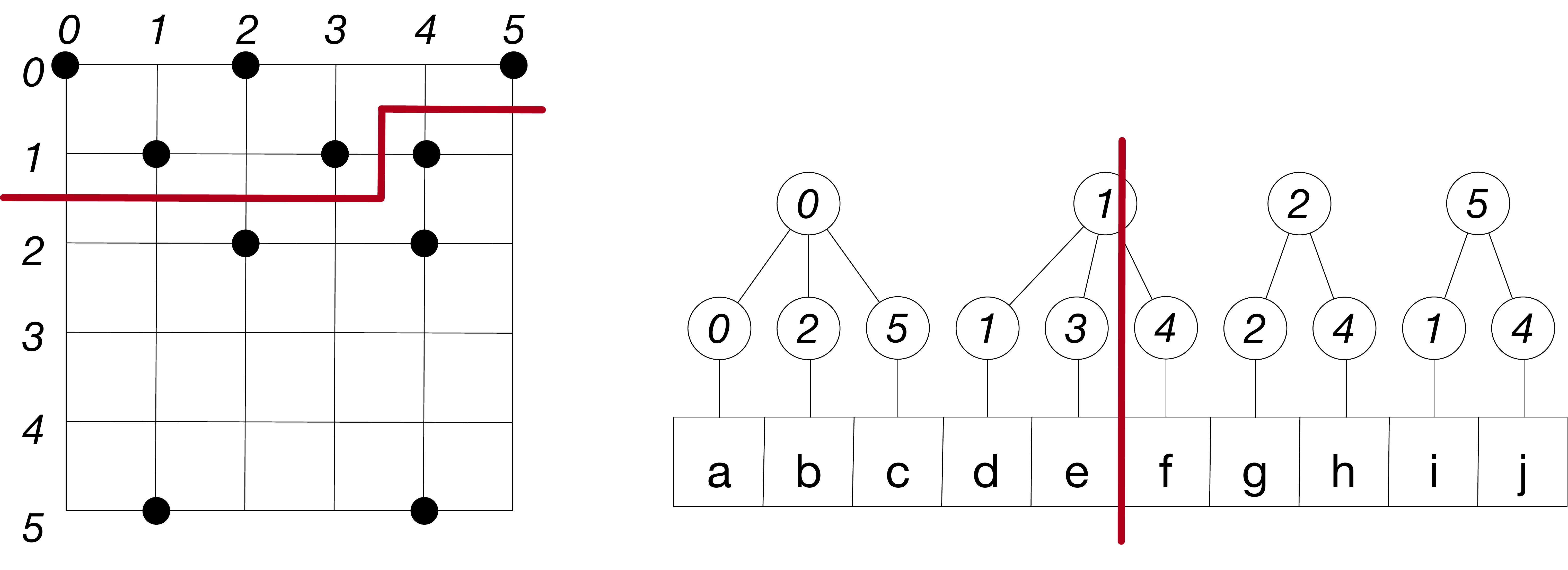}
    \caption {
      Tiling the position space of the column coordinates.
      \label{fig:position-cut}
    }
  \end{minipage}
  \begin{minipage}{\linewidth}
    \centering
    \includegraphics[width=\linewidth]{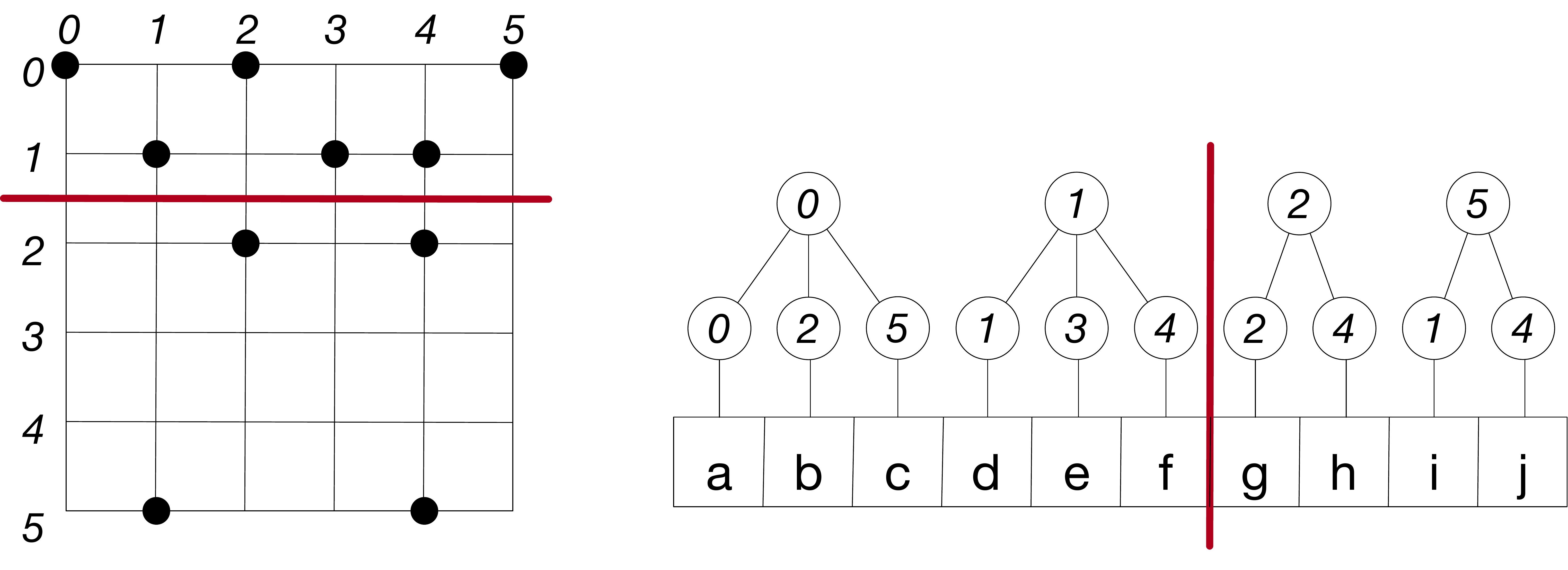}
    \caption {
      Tiling the position space of the row coordinates.
      \label{fig:row-cut}
    }
  \end{minipage}
\end{figure}

We can tile the coordinate space or the position space of a sparse
tensor algebra expression's loop nest.  Tiling its coordinate
space is conceptually straightforward and similar to how tiling
works for dense loops.  We simply split the dimensions of the
sparse space into coordinate sets that are assigned to separate
tiles.  The challenge with tiling a sparse iteration space is that
the tiles may not have the same number of nonzeros and 
will therefore have different sizes in memory.  This means that a
potentially expensive search is required to determine the starting
location of each tile when iterating over them.  Furthermore, 
varying size blocks lead to different amounts of computation and
thus load-imbalanced parallel execution.  \figref{coordinate-cut}
demonstrates the issue: although the cuts result in tiles with
equal area, each tile contains very different numbers of points
in its iteration space. We see this clearly when projecting
the cut to the coordinate hierarchy data structure of $B$, where
the number of values in each tile substantially differs.

Tiling the position space of a sparse tensor algebra expression
opens up new and exciting opportunities.  The key
property of position spaces are that they, in contrast to
coordinate spaces, are dense and contiguous.  The tiles therefore
have the same size and can easily be located.  This lets us create
statically load-balanced parallel code and makes it possible to
generate code that is tuned for GPUs, which tend to benefit from
more regular execution than CPUs.  \figref{position-cut} shows the
effect of tiling in the position space on the sparse coordinate
space of the \spmv example and on its data structures.  The
position tiles, in contrast to coordinate space tiles, are
irregular in the coordinate space, but lead to equal-sized tiles
in the position space.  However, 
tiling in the position space introduces book-keeping code to track
the coordinates above the cut (e.g., what row we are on in the
\spmv example) and can lead to conflicting writes when a row spans
two tiles.  Furthermore, we can only cut the position space with
respect to the data structure of one of the operands of the
expression.  The resulting code, as we will see
in~\secref{compiling}, iterates over this operand, computes
coordinates, and then finds the position of those coordinates in
the other operands.  Despite these drawbacks, we find that
position tiles are often crucial, especially on inflexible compute
platforms such as GPUs.  For example, they let us recreate
important optimized GPU codes from the literature, while
generalizing to many tensor algebra expressions not previously
studied.

Finally, we can generalize position tiling to apply to any
dimension of the iteration space, which corresponds to different
levels of coordinate hierarchy in the data structure whose positions we
are tiling.  This lets us create tiles with a fixed number of
coordinates in any dimensions.  For example, we can tile a \spmv 
operation in the position space of the rows of matrix $B$.  This
creates tiles that have the same number of rows; however, the
rows themselves may have different numbers of nonzeros.
\figref{row-cut} shows a position cut in the row dimension of the
iteration space, which corresponds to the first level of the
coordinate hierarchy.  The cut evenly divides the rows, whereas
the position cut in the lowest level of the hierarchy evenly
divided the nonzero coordinates.  The benefit of a cut in a higher
level of the hierarchy is that the cut does not divide a row in
two.  This results in less book-keeping code and avoids
synchronization as the computation of two tiles do not write to
the same locations in the result.

\section{Intermediate Representation}
\label{sec:ir}
In order to carry out the iteration space transformations 
in this paper, we extend the iteration graph
intermediate representation of Kjolstad et al.~\cite{kjolstad2017}
with the concepts of derived index variables, position index
variables, and parallel index variables.  Iteration graphs
describe the lexicographically ordered sparse iteration space that
results from iterating over coordinate hierarchy
tensor data structures, and transformations on the graphs
transform that space.  Combined with iteration graph code
generation, described in the next section, we can create loop
nests that iterate over the space by coiterating over coordinate
hierarchies.

\subsection{Iteration Graph Background}

We can symbolically describe an iteration space by a
lexicographical ordering of index variables that represent
dimensions, as in the polyhedral model.  We extend this to a tree
of such variables, where the concatenated ranges of the variables
of each tree level together encode a dimension.  This is analogous
to an imperfect loop nest where an outer loop contains two or more
sequenced loops.  These index variable trees describe the full, or
dense, iteration space of a tensor algebra expression.

\begin{figure}[]
	\centering
  \includegraphics[width=1\linewidth]{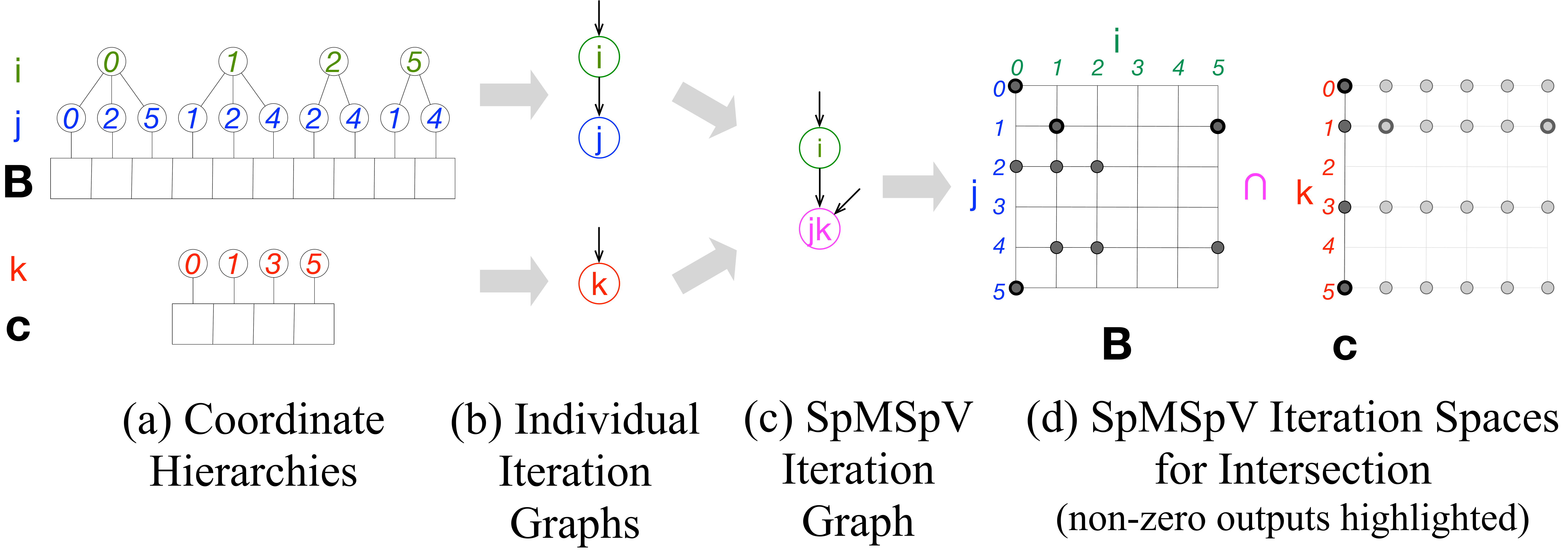}  
  \caption {
    Coordinate hierarchies and iteration spaces of a sparse matrix
    B and sparse tensor c are multiplied (SpMSpV). 
		\label{fig:iteration-graphs}
	}
\end{figure}

Sparse iteration graphs extend iteration variable trees with paths
through index variable nodes that each represent a coordinate tree
hierarchy that encode a subset of the iteration space.  In tensor
algebra these coordinate hierarchies, as we saw in the previous
section, come from the data structures that encode the nonzero
values of tensors.  \figref{iteration-graphs}a shows two abstract
coordinate hierarchies for the matrix-vector multiplication
example.  The expression has two index variables that we choose to
lexicographically order $i$ before $j$.
\figref{iteration-graphs}b shows the two symbolic paths induced by
these hierarchies.  The matrix $B$ enumerates both $i$ and $j$
coordinates and therefore its path go through both variables,
where the vector $c$ only enumerates $j$ coordinates.  Finally,
\figref{iteration-graphs}c shows the iteration graph that contains
both paths, since the whole expression must coiterate over both
coordinate hierarchies.  The graph is annotated with an
intersection operation between the two paths incoming on $j$.
Thus, the iteration space of $j$ is the intersection of each row
of $B$ and the vector $c$.  It is an intersection because
multiplying any variable by zero yields a zero.  It is therefore
sufficient to iterate over those coordinates where both operands
have a value in order to compute the output nonzeros.  Conversely,
a tensor addition would induce a union between incoming paths.

\subsection{Derived Index Variables and Provenance Graphs}
\label{derived-index-variables}

\begin{figure}[]
  \centering
  \small{
	  \begin{tabular}{l c c c}
	  & Iteration & Nested    & Unrolled  \\
	  & Graph     & Iteration & Iteration \\
	  Original &
	  \includegraphics[width=.04\linewidth]{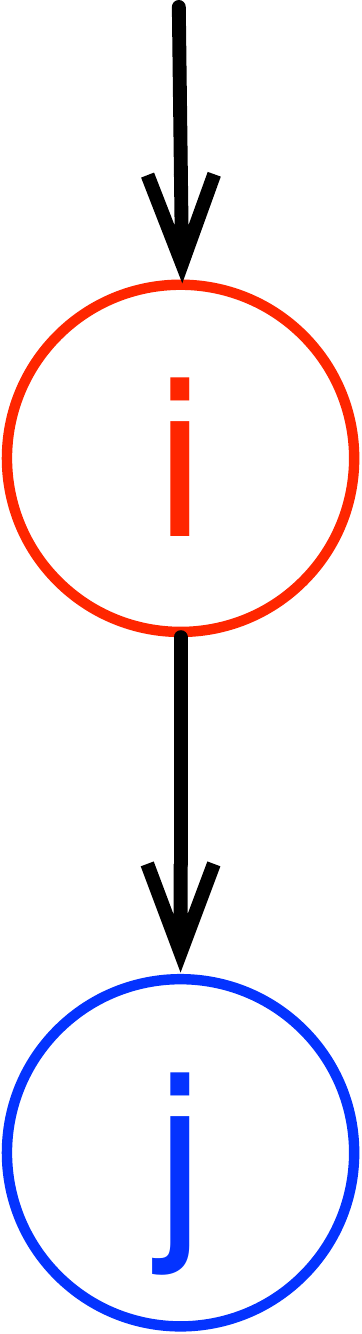}  &
	  \includegraphics[width=.28\linewidth]{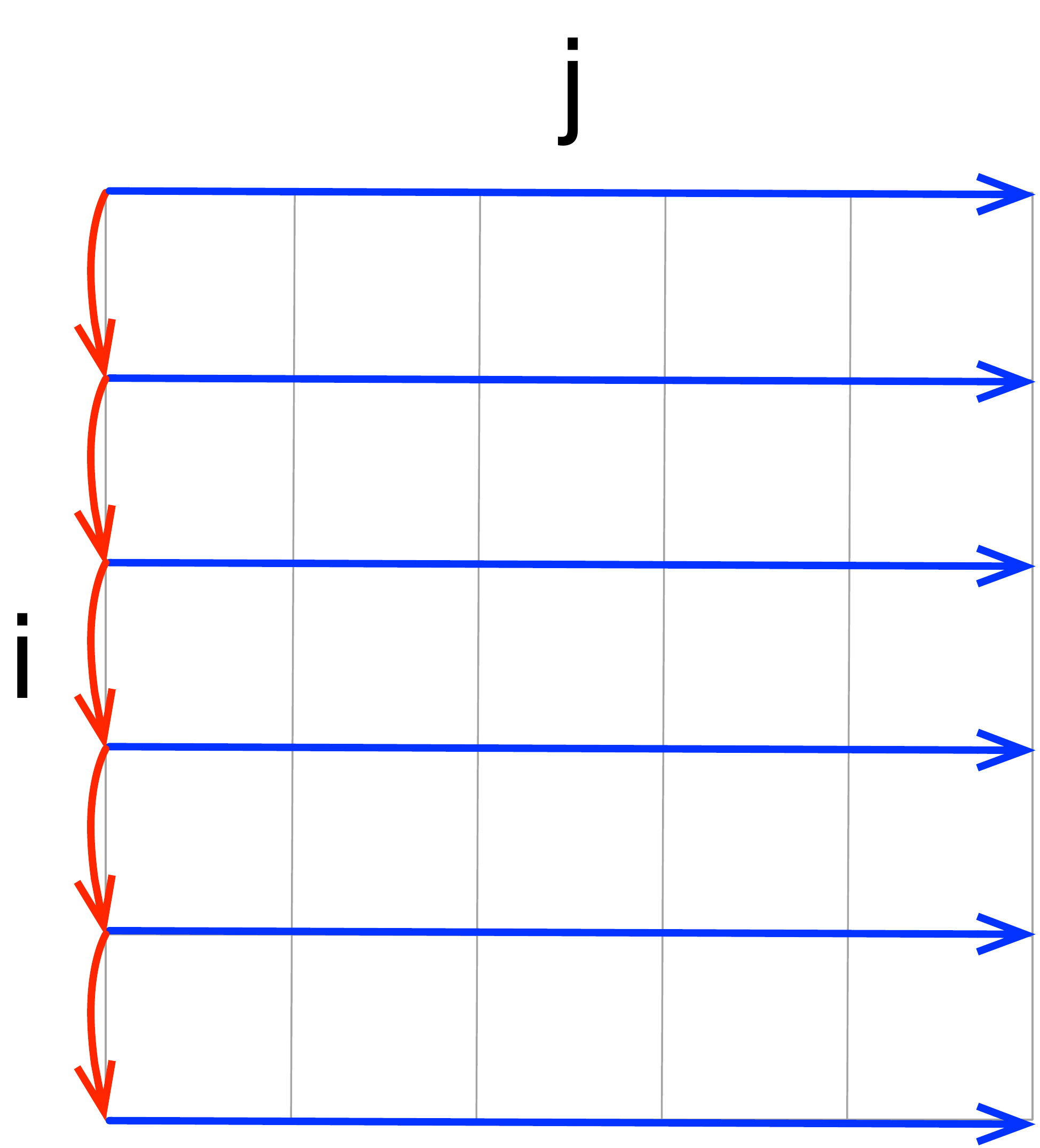}  &
	  \includegraphics[width=.28\linewidth]{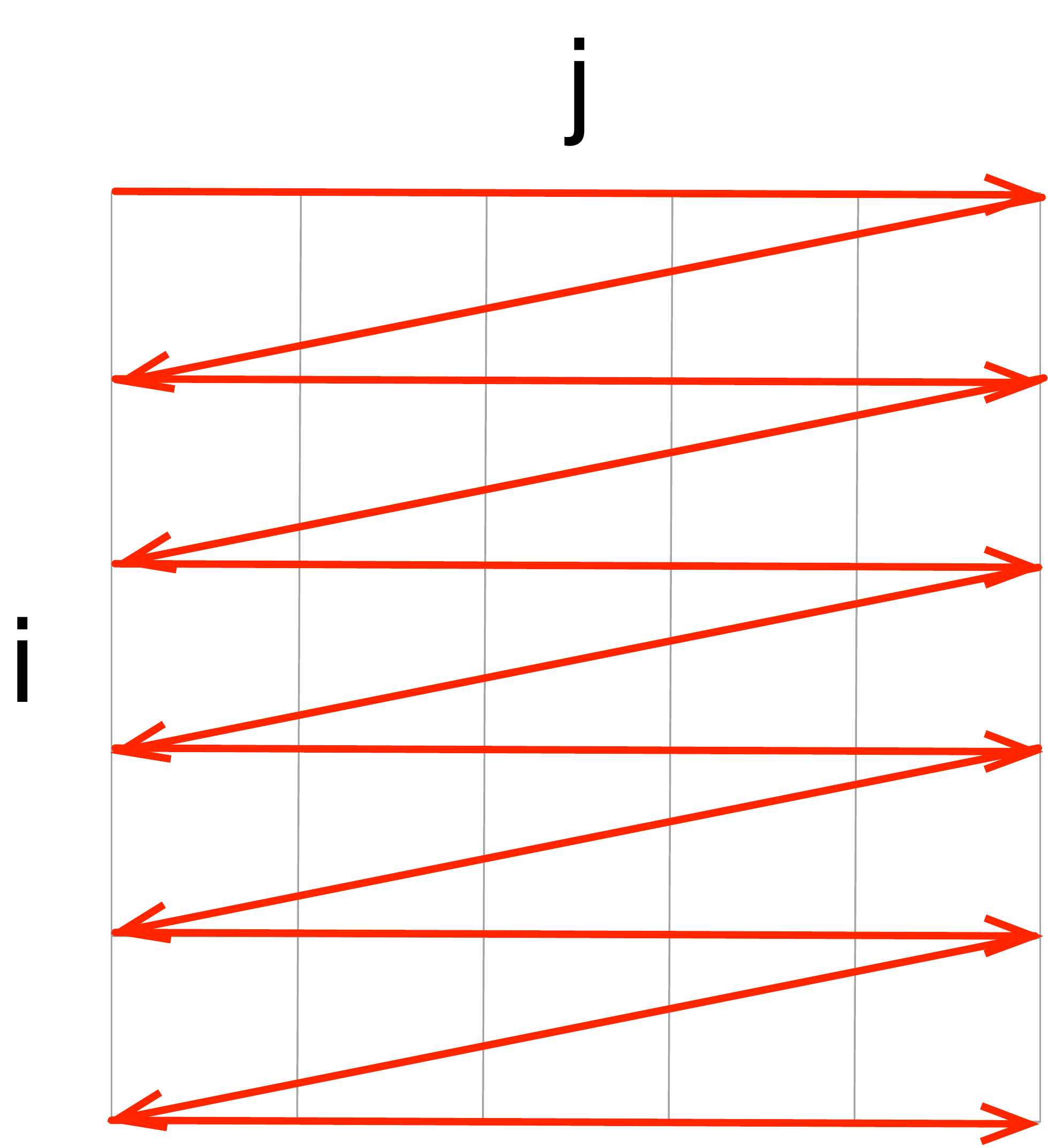} \\
	  Tiled &
	  \includegraphics[width=.035\linewidth]{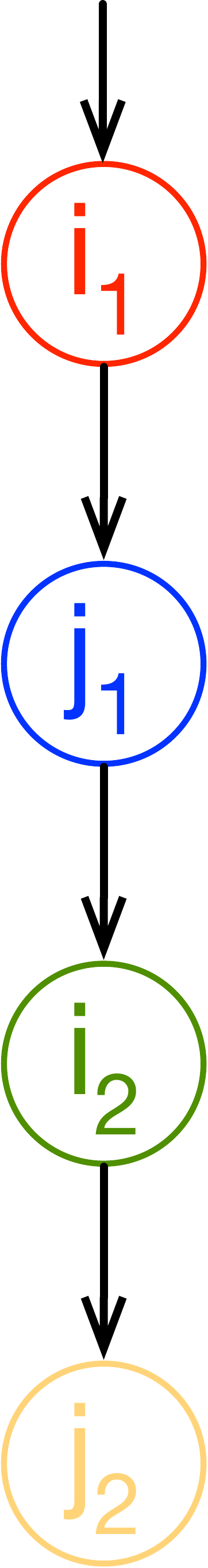}  &
	  \includegraphics[width=.28\linewidth]{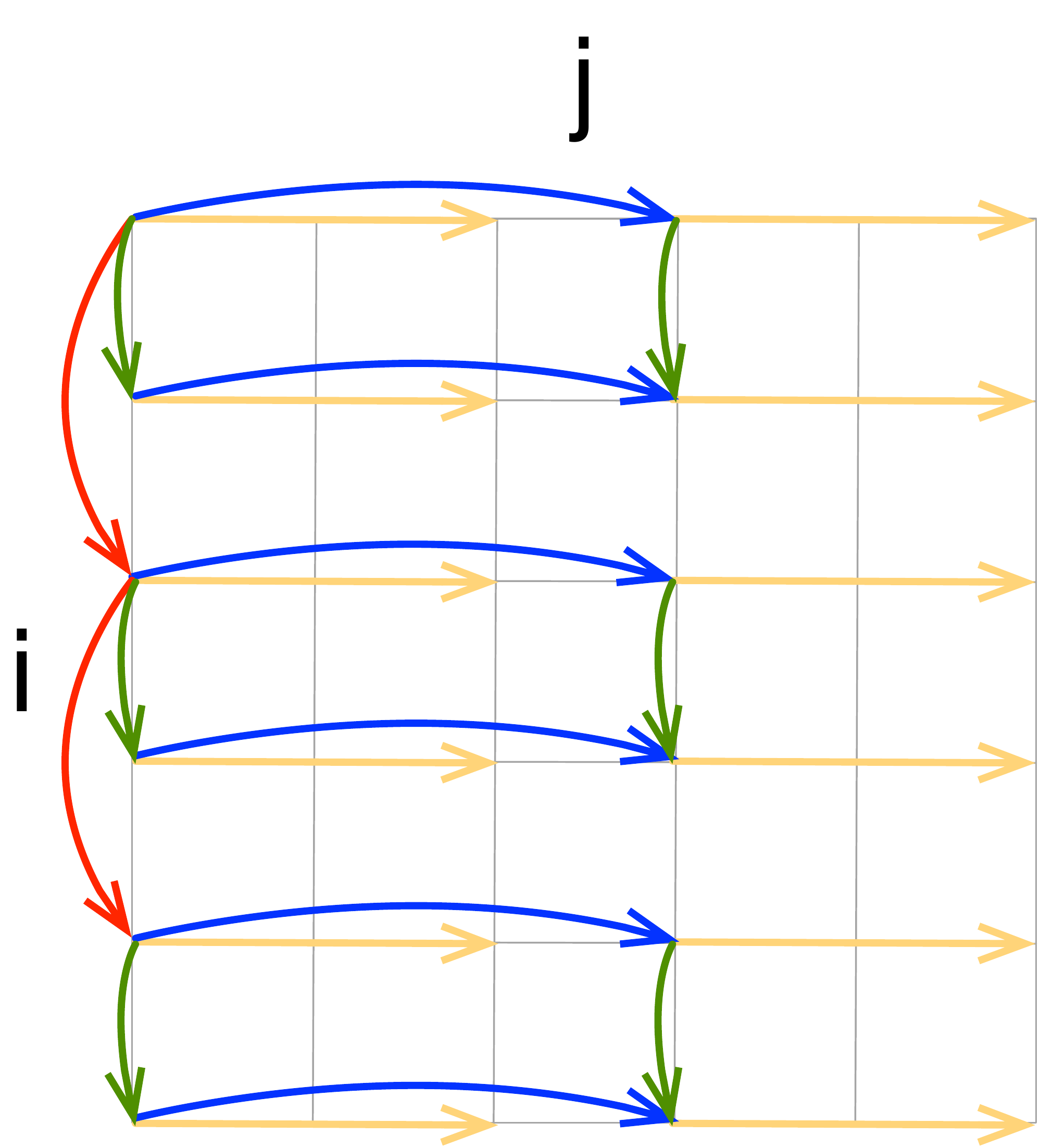}  &
	  \includegraphics[width=.28\linewidth]{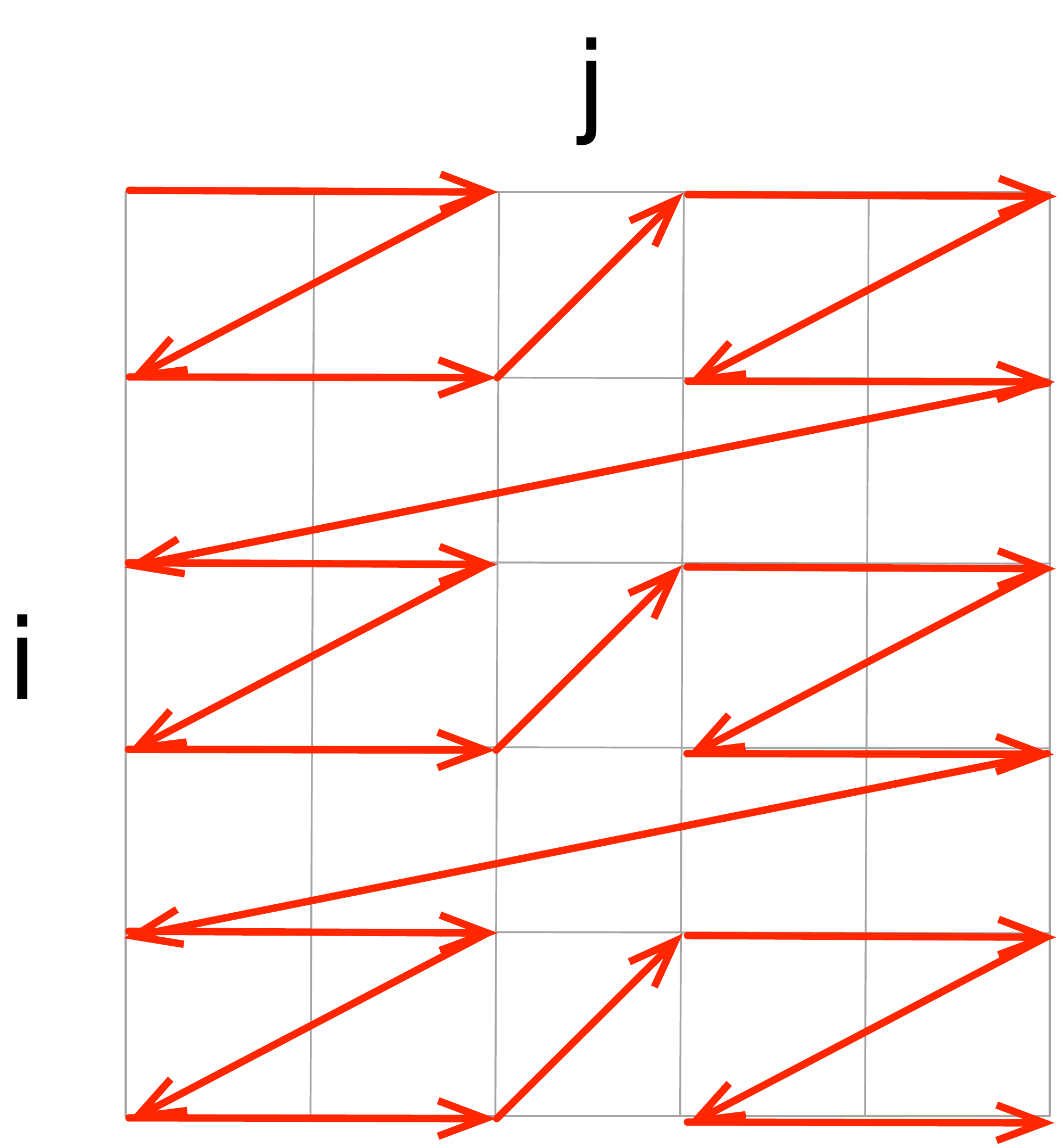}  \\
	  \end{tabular}
  }
  \caption {
    Iteration graphs, nested iteration orderings, and unrolled
    iteration orderings are shown for an untiled row-major
    iteration (top) and a tiled iteration (bottom).
    \label{fig:tiledigraph}
  }
\end{figure}

In this paper, we extend iteration graphs with the concept of
\textbf{derived index variables}.  These are new dimensions that
are added to an expression's iteration space by the split, fuse,
and position iteration space transformations (described in
\secref{transformations}) and can be in either coordinate space or
position space.  \figref{tiledigraph} shows an iteration space
before and after it has been tiled by splitting and reordering the
index variables.  The tiling increases the dimensionality from two
dimensions to four.  Since we cannot cleanly visualize a
four-dimensional space, however, we visualize the tiled iteration
space in terms of its iteration order when projected onto the
original iteration space.  The nested iteration diagrams show the
iteration of each index variable as differently colored arrows.
For example, in the original iteration space, the iteration
proceeds along the first blue arrow before it moves along the
first red arrow to the next row.  The unrolled iteration space
shows the order of the overall iteration.  Mapping between
iteration spaces is important for code generation, since emitted
loops iterate in the transformed space, while data structures must
be accessed by coordinates in the original space.

\begin{figure}
  \centering
  \includegraphics[width=0.9\linewidth]{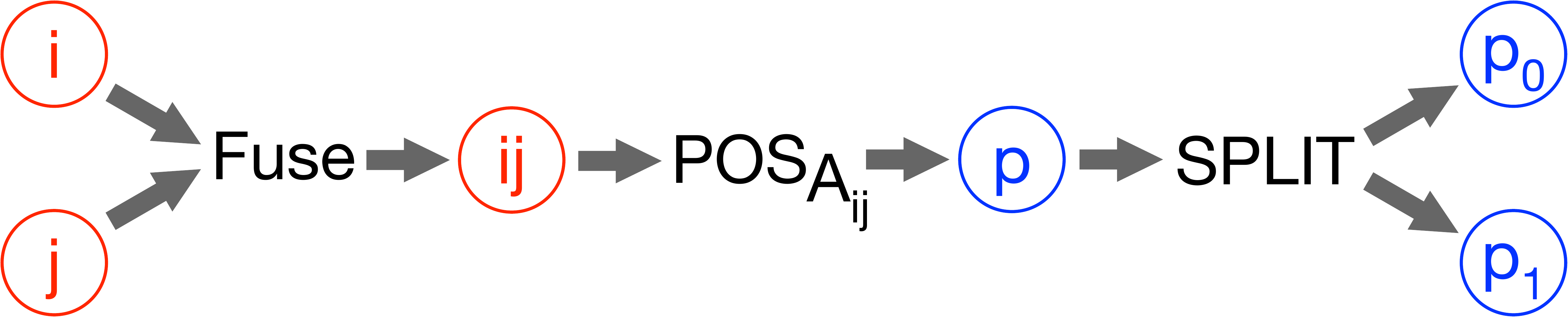}
  \caption{ 
    Index variable provenance graph after parallelizing SpMV in
    the CSR matrix's position space.  The graph maps derived index
    variables back to the index variables they derived from.
    Blue variables are in position space.
    \label{fig:provenance-graph}
  }
\end{figure}

Index variable \textbf{provenance graphs} track the history of
derived index variables back to the original index variables.
Provenance graphs let us map between the transformed iteration
space and the original space, so that the code generator can
compute coordinates in the original space as needed to index into
tensors data structures.  \figref{provenance-graph} shows the
provenance graph after creating a load-balanced SpMV kernel by
tiling the expression in the position space.  The derived index
variables that represent the dimensions of the final iteration
space, \code{p0} and \code{p1}, are tracked back to the index
variables in the original space through the transformations they
went through (transformations are described in the next section).

\section{Transformations}
\label{sec:transformations}
In this section, we will describe a transformation framework for
sparse, dense, and mixed sparse/dense iteration spaces.  The
framework operates on the iteration graph intermediate
representation discussed in the previous section.  These
transformations let us control the order of computation, so that
we can optimize data access locality and parallelism.

\begin{figure}
  \center
  \begin{minipage}[b]{0.325\linewidth}
    \includegraphics[width=0.95\linewidth]{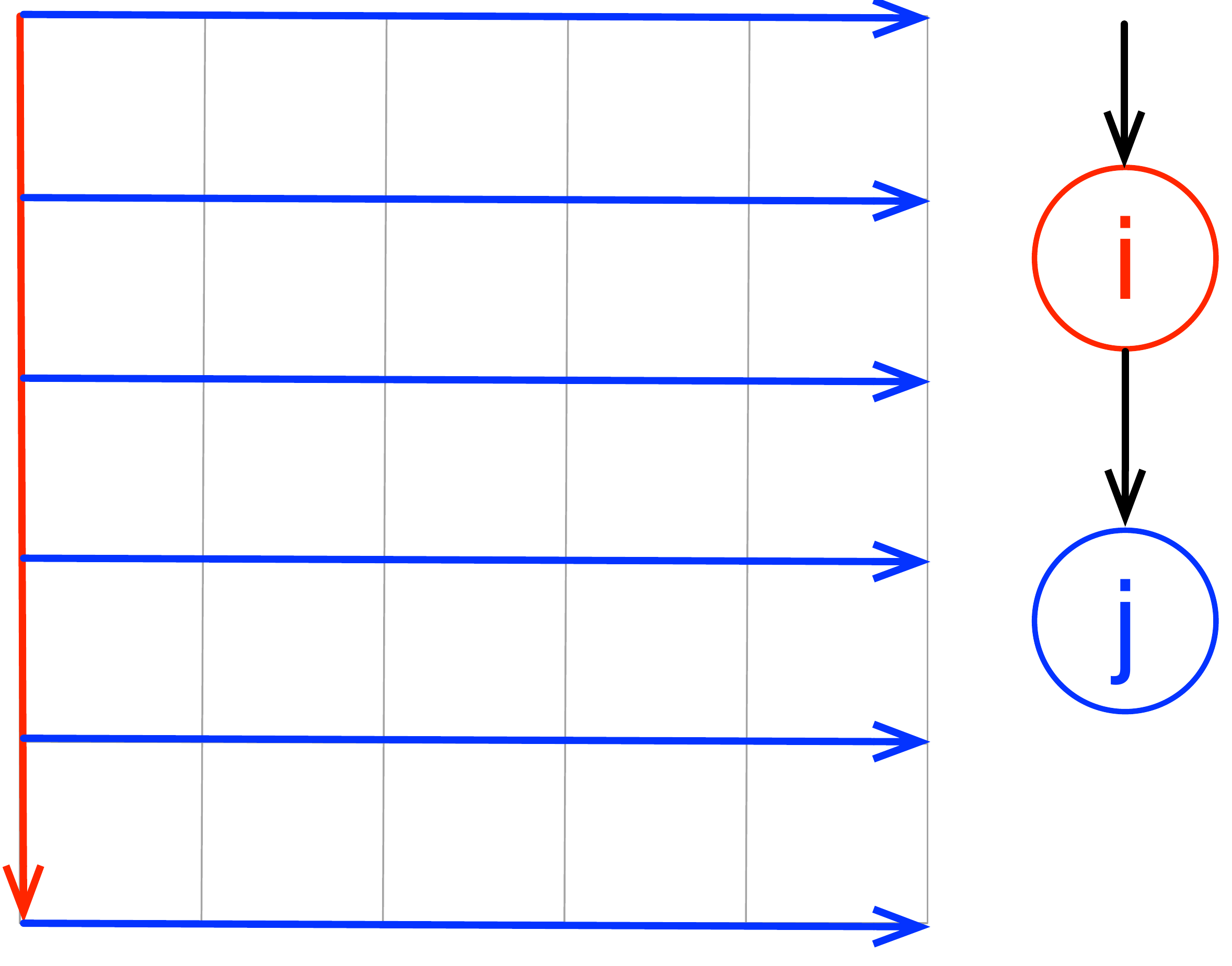}
    \subcaption{
      Original
      \vspace{1mm}
      \label{fig:space-transformations-original}
    }
  \end{minipage}
  \begin{minipage}{\linewidth}
    \begin{minipage}[b]{0.325\linewidth}
      \includegraphics[width=0.95\linewidth]{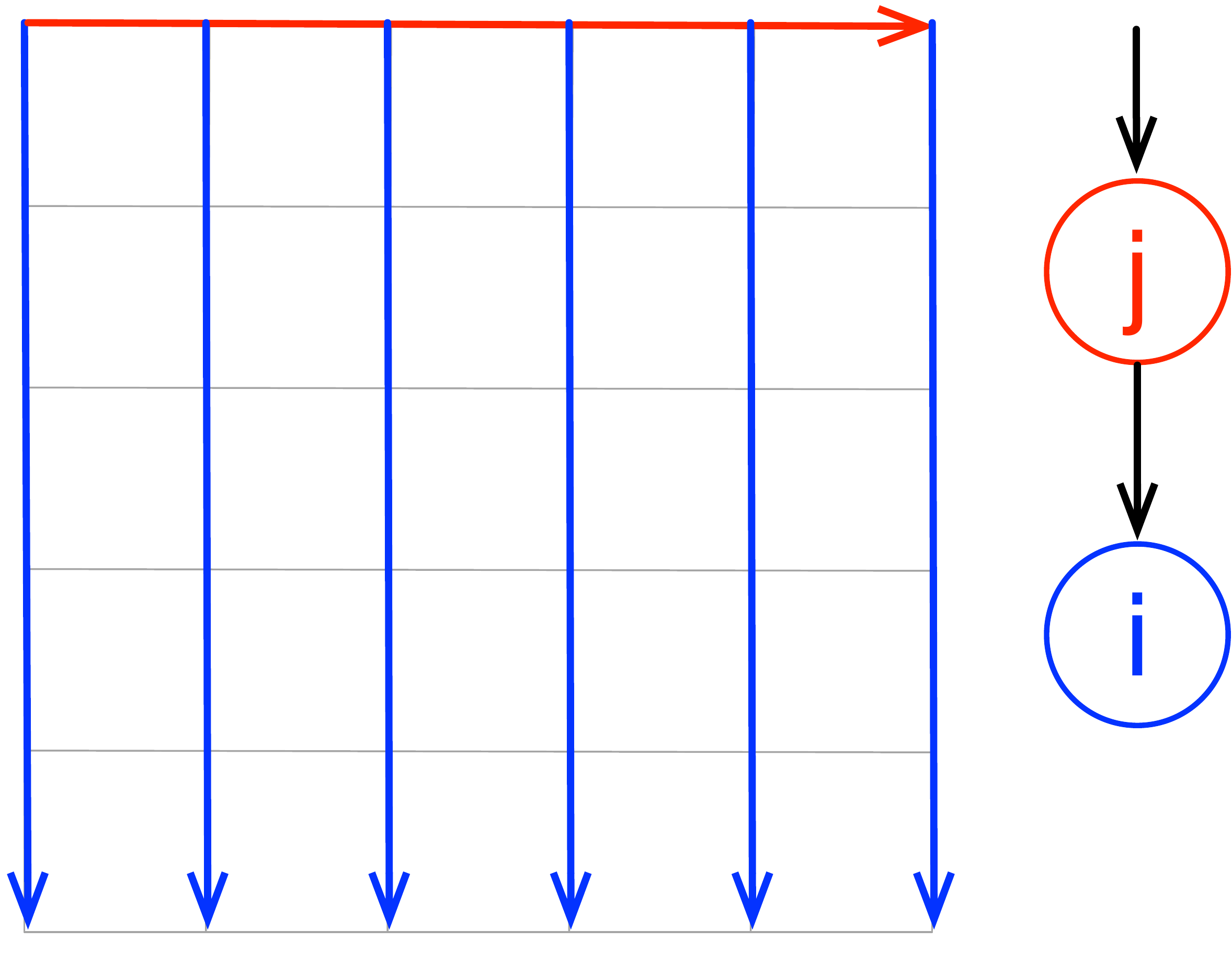}
      \subcaption{
        reorder(i,j)
        \label{fig:space-transformations-reorder}
      }
    \end{minipage}
    \begin{minipage}[b]{0.325\linewidth}
      \center
      \includegraphics[width=0.95\linewidth]{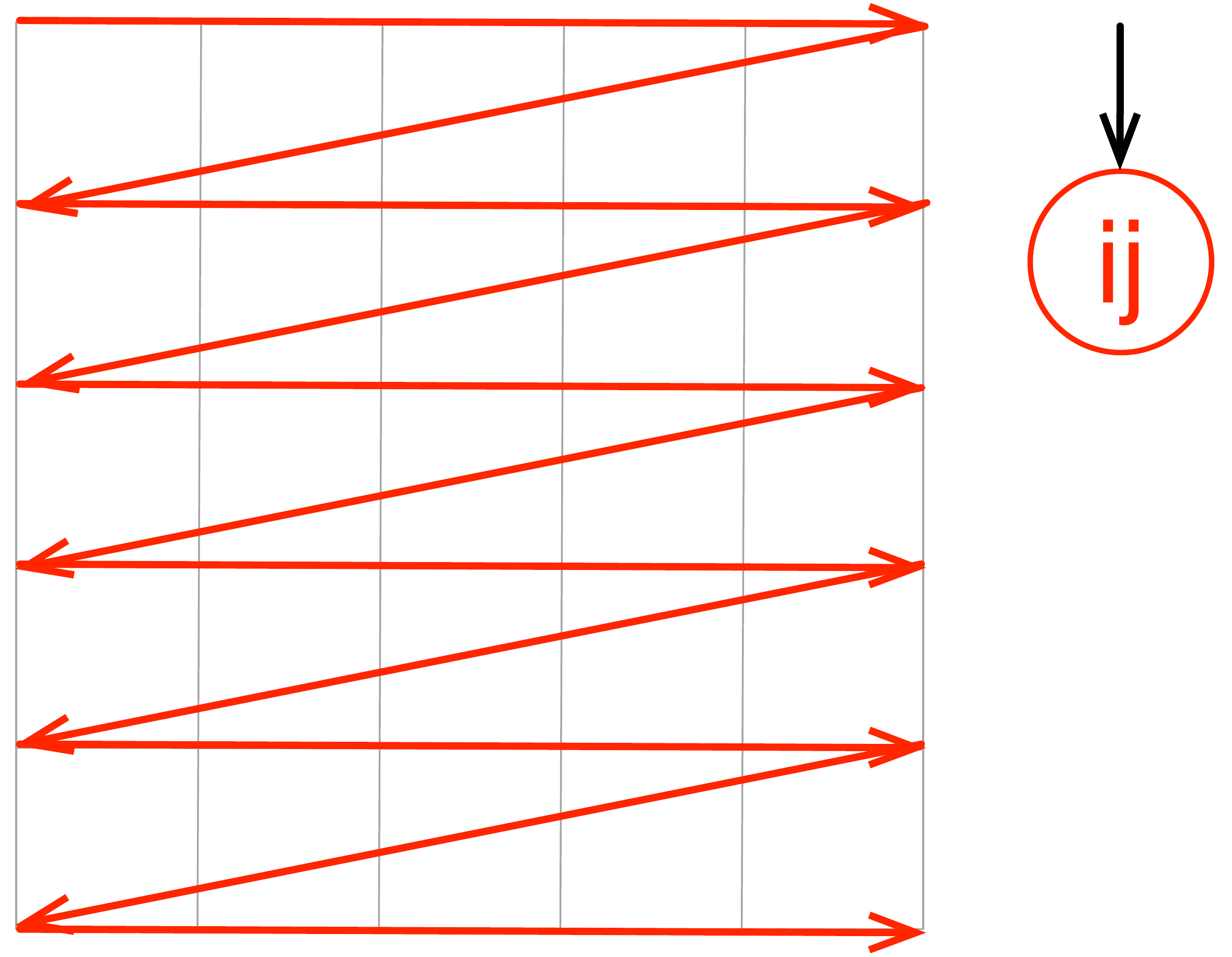}
      \subcaption{
        fuse(i,j,ij)
        \label{fig:space-transformations-fuse}
      }
    \end{minipage}
    \begin{minipage}[b]{0.325\linewidth}
      \includegraphics[width=0.95\linewidth]{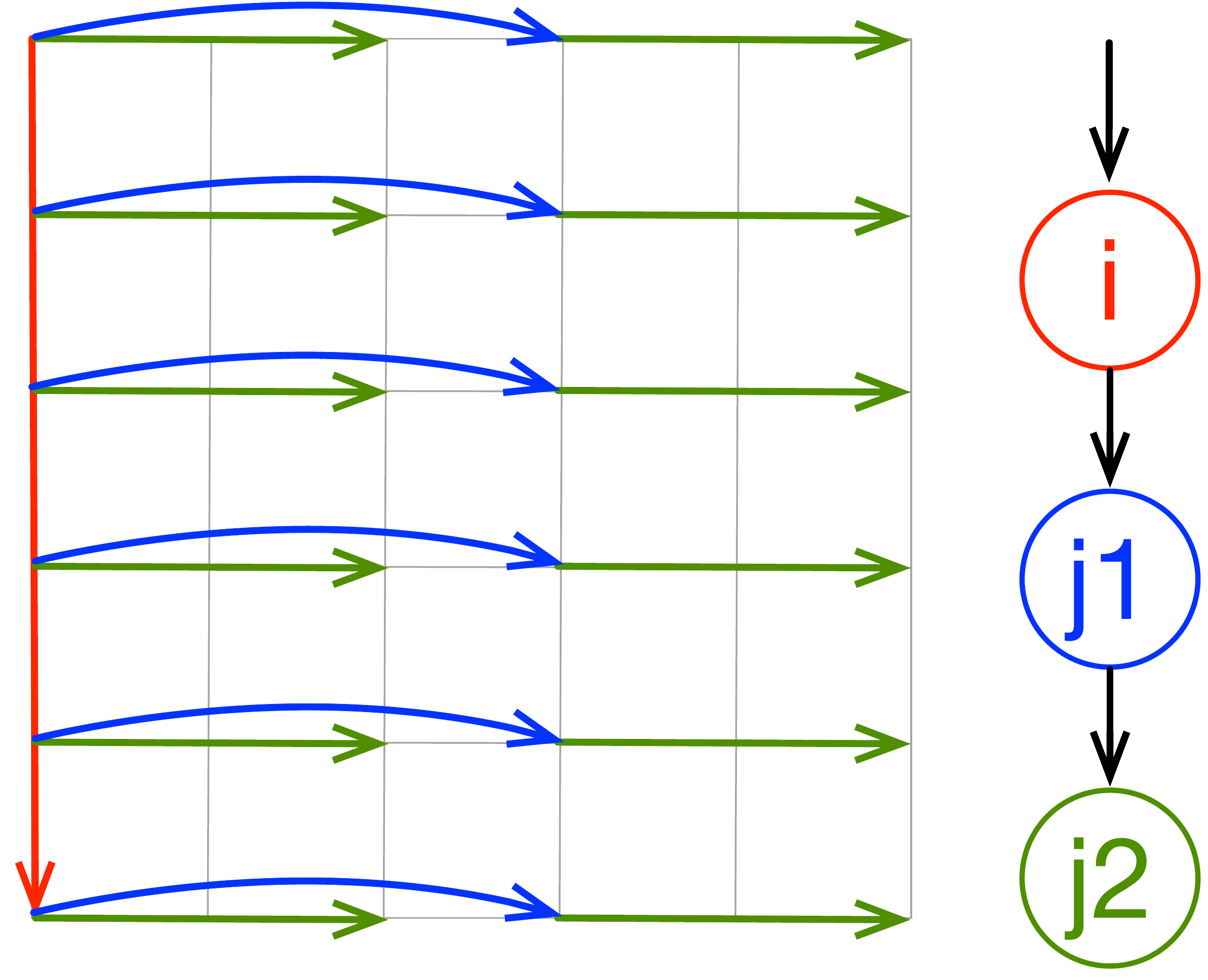}
      \subcaption{
        split(j,j1,j2)
        \label{fig:space-transformations-split}
      }
    \end{minipage}
  \end{minipage}
  \caption {
    An original row-major nested iteration of a two-dimensional
    coordinate iteration space is shown alongside various
    transformed nested iterations caused by different iteration
    space transformations.
    \label{fig:space-transformations}
  }
\end{figure}

The transformations in this paper---\code{coord}, \code{pos},
\code{reorder}, \code{fuse}, \code{split}, \code{divide} and
\code{parallelize}---provide a comprehensive framework for
controlling iteration order through sparse iteration spaces.
\figref{space-transformations} shows the effect of the reorder,
fuse and split transformations on an $i,j$ iteration space.
Although they are here shown separately, transformations are
typically used together, with some adding or removing iteration
space dimensions that other transformations reorder or tag for
parallel execution.  All transformations apply to index variables
in both the coordinate space and the position space and the
\code{coord} and \code{pos} transformations transition index
variables between these spaces.

Key to our approach is that the transformations operate on the
iteration graph intermediate representation before sparse code is
generated.  This representation makes it possible to reason about
sparse iteration spaces algebraically without the need for
sophisticated dependency and control flow analysis of sparse code,
which may contain while loops, conditionals and indirect accesses.
The sparse code is then introduced when iteration graphs are
lowered to code, as described in the next section.

\paragraph{Coordinate and Position Transformations}

The \code{coord} and \code{pos} transformations
create new index variables in the coordinate and position spaces
from index variables in the position and coordinate spaces
respectively.  Specifically, the \code{pos} transformation takes a
index variable in the coordinate space and replaces it with a new
derived index variable that operates over the same iteration
range, but with respect to one input's position space.  The
\code{coord} transformation, on the other hand, takes an index
variable in position space and replaces it with a new derived
dimensional iterator that iterates over the corresponding
iteration range in the coordinate iteration space.

\paragraph{Reorder}

The \code{reorder} transformation swaps two directly nested index
variables in an iteration graph.  This changes the order of
iteration through the space and the order of tensor accesses.  The
precondition of a reorder transformation is that it must not hoist
a tensor operation outside a reduction that it does not distribute
over.  \figref{space-transformations-reorder} shows the effect of
the reorder transformation on a two-dimensional iteration space,
in terms of iteration order on the original space. Whereas the
original space was iterated through in row-major order, the
reordering creates a new space that is equivalent to iterating
through the original space in column-major order.

\paragraph{Fuse}

The \code{fuse} transformation collapses two directly nested index
variables.  It results in a new fused index variable that iterates
over the product of the coordinates of the fused index variables.
This transformation by itself does not change iteration order, but
facilitates other transformations such as iterating over the
position space of several variables and distributing a
multi-dimensional loop nest across a thread array on GPUs.
\figref{space-transformations-fuse} shows the iteration order of
the fused space in terms of the original space.  The fused space
is a one-dimensional space that is equivalent to iterating over
the original space in a linearized row-major order.

\paragraph{Split}

The \code{split} transformation splits (strip-mines) an index
variable into two nested index variables, where the size of the
inner index variable is constant.  The size of the outer index
variable is the size of the original index variable divided by the
size of the inner index variable, and the product of the new index
variables sizes therefore equals the size of the original index
variable.  Note that in the generated code, when the size of the
inner index variable does not perfectly divide the original index
variable, a \textit{tail strategy} is employed such as emitting a variable
sized loop that handles remaining iterations.
\figref{space-transformations-split} shows the effect of the split
transformation as an iteration order over the original space.  The
split creates a new inner loop that iterates over two iteration
space points for each iteration of the outer loops.

\paragraph{Divide}

The \code{divide} transformation splits one index variable into
two nested index variables, where the size of the \emph{outer}
index variable is constant.  The size of the inner index variable
is thus the size of the original index variable divided by the
size of the outer index variable.  The \code{divide}
transformation is important in sparse codes because locating the
starting point of a tile can require an $O(n)$ or $O(\log (n))$
search.  Therefore, if we want to parallelize a blocked
loop, then we want a fixed number of blocks and not a number
proportional to the tensor size.

\paragraph{Parallelize, Bound and Unroll}

The parallelize, unroll and bound transformations apply to only
one index variable and tag it with information telling the code
generation machinery how to lower it to code.  The \code{parallelize}
transformation tags an index variable for parallel execution.  The
transformation takes as an argument the type of parallel hardware
to execute on.  The set of parallel hardware is extensible and our
current code generation algorithm supports SIMD vector units, CPU
threads, GPU thread blocks, GPU warps, and individual GPU threads.
Parallelizing the iteration over an index variable changes the iteration
order of the loop, and therefore requires reductions inside the
iteration space described by the index variable's sub-tree in the
iteration graph to be associative.  Furthermore, if the
computation uses a reduction strategy that does not preserve the
order, such as atomic instructions, then the reductions must also
be commutative.  The \code{unroll} transformation tags an index
variable to result in an unrolled loop with a given unroll factor.
This reduces the amount of control flow logic at the cost of
increased code size.  Finally, the \code{bound} transformation
fixes the range of an index variable, which lets the code
generator insert constants into the code and enables other
transformations that require fixed size loops, such as
vectorization.

\section{Code Generation}
\label{sec:compiling}
We extend the sparse tensor algebra code generator described by
Kjolstad et al.~\cite{kjolstad2017} to support
iteration graphs with derived index variables.  This is sufficient
to lower transformed iteration spaces to efficient code and to
generate parallel and GPU code.

The existing code generator operates on iteration graphs and
generates loops to iterate over each index variable in turn,
nested inside the loops generated for index variables above in the
tree.  For each index variable, one or more loops are generated to
either iterate over a full dimension (a dense loop) or to
coiterate over levels of one or more coordinate hierarchy data
structures.  Coiteration code is generated using a construct
called a \textit{merge lattice} that enumerates the intersections
that must be covered to iterate over the sparse domain of the
dimension.  This may result in a single for loop, a single while
loop, or multiple while loops.

To generate code for transformed sparse iteration spaces, we
extended this code generator to
\begin{enumerate}

  \item determine derived index variable bounds, generate loops
    over derived index variables, and generate iteration guards to
    implement a tail strategy,

  \item generate code to recover coordinates in the original
    iteration space from coordinates in the transformed iteration
    space, and

  \item generate SIMD vectorized, OpenMP CPU and CUDA GPU code
    with reductions.

\end{enumerate}

\begin{figure}
  \centering
  \includegraphics[width=\linewidth]{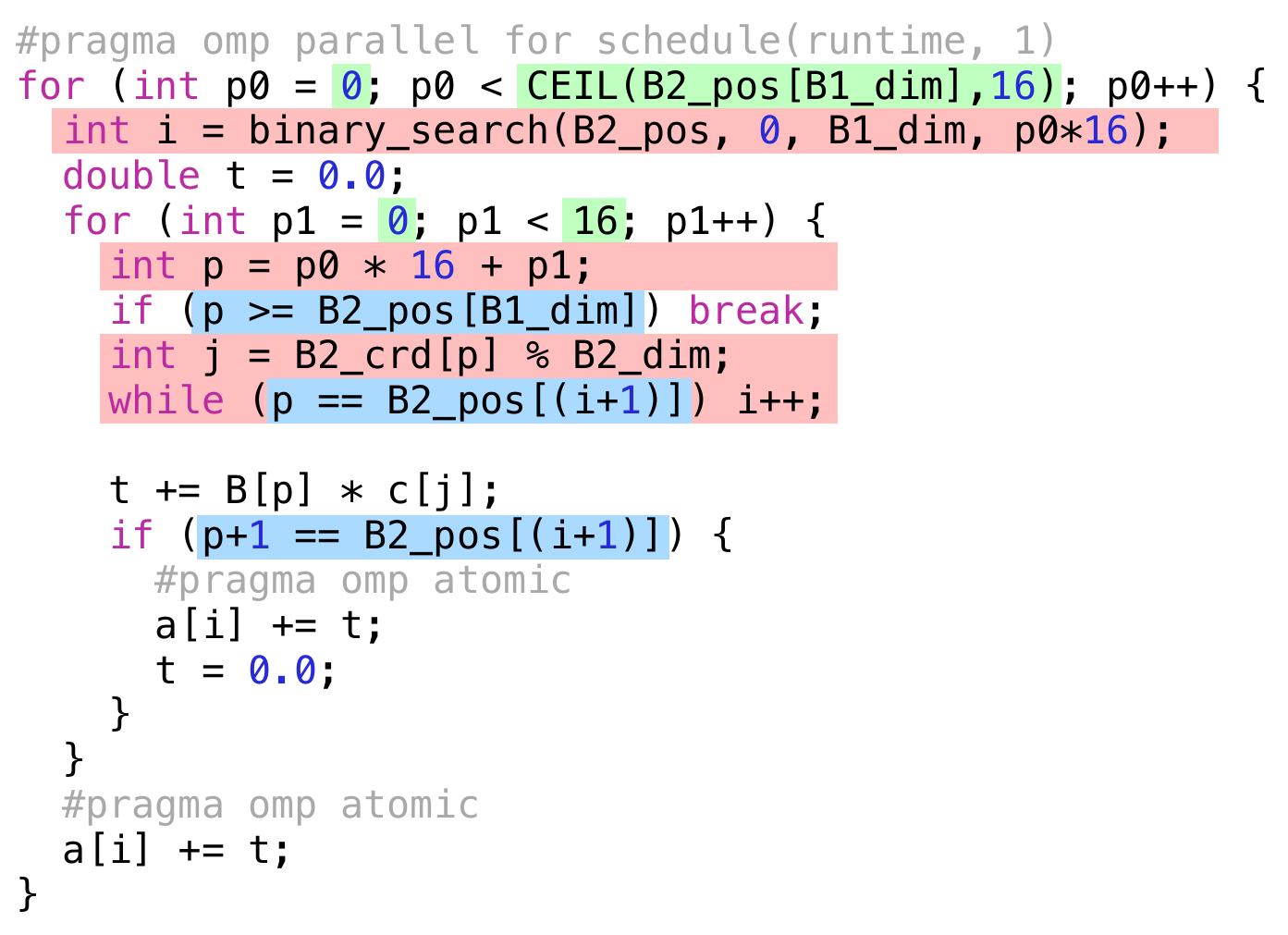}
  \caption {
    Generated code for parallel
    sparse matrix-vector multiply.  Red code recovers index
    variables, green code shows iteration bounds of derived index
    variables, and blue code depicts iteration guards.
    \label{fig:generated-code}
  }
\end{figure}

\figref{generated-code} provides an example of a generated \spmv
implementation that highlights derived loop bounds (green),
iteration guards (blue), and index variable recovery (red).  The
following sections describe each of these extensions.  

\subsection{Derived Iteration Code Generation}

To generate loops that iterate, or coiterate, over derived index
variables, the code generator must first compute their iteration
domains.  These domains are computed by an iteration domain
propagation algorithm, and affect the bounds of generated loops
(green in \figref{generated-code}).  In cases where a fixed range
index variable was split of from another index variable, and its
size does not evenly divide the original variable, the code
generation algorithm also generates an iteration guard tail
strategy (blue in \figref{generated-code}).\footnote{Note that we
show the iteration guards inside the loop for readability, but in
our implementation we clone the loop and apply the iteration guard
outside to determine which loop to enter.}

To determine the iteration domain of derived index variables, we
propagate bounds through the index variable provenance graph
(see~\figref{provenance-graph}).  We have defined propagation
rules for each transformation and calculating the iteration domain
of the derived index variables involves applying the propagation
rules to each arrow in turn, from the original index variables to
the derived index variables.

\subsection{Coordinate Recovery}

In a transformed iteration space dimensions are represented by
derived index variables and the generated loops iterate over their
coordinates.  The coordinate hierarchy data structures of the
tensors, however, contain and are accessed by coordinates in the
original iteration space.  The code generator must therefore emit
code to map between these iteration spaces and we call this
coordinate recovery.

It may be necessary to recover original or derived coordinates.
Recovering original coordinates is required when these will be
used to index into tensor data structures, which are stored in the
original coordinate space.  It is necessary to recover derived
coordinates, on the other hand, when a coordinate in the original
coordinate space is loaded from a coordinate hierarchy, but a
coordinate in the derived space is needed to determine iteration
guard exit conditions.

\begin{figure}
  \centering
  \includegraphics[width=0.9\linewidth]{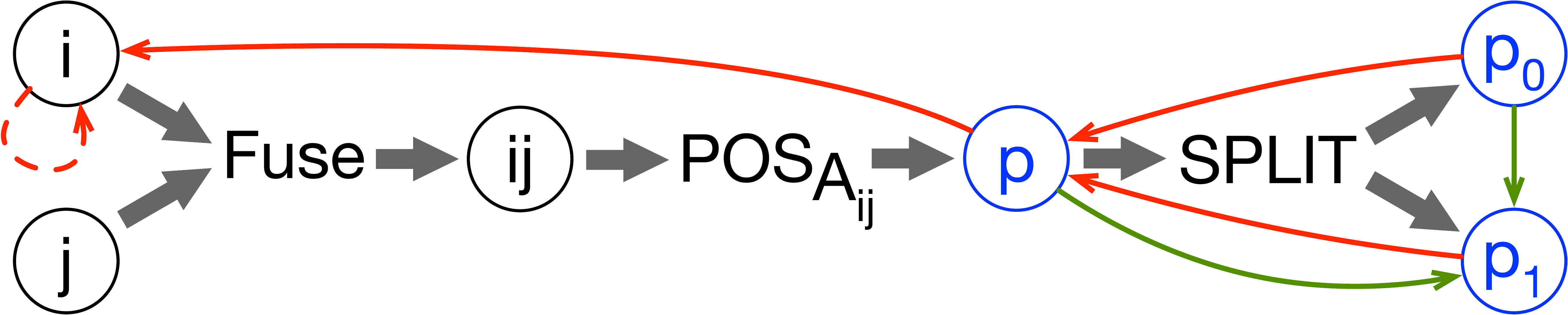}
  \caption{
    An index variable provenance graph annotated with arrows that
    depict different ways that an unknown index variable's
    coordinates can be recovered from known index variables.  Red
    arrows depict original coordinate recovery and green arrows
    derived coordinate recovery.
    \label{fig:compiling-recovery} 
  }
\end{figure}

The code generator  defines two functions on the provenance graph
to map coordinates between original and derived index variables,
and vice versa. These are:
\begin{description}

  \item [recover\_original] which computes the coordinate of an
    index variable from its derived index variables in a
    provenance graph (red arrows in~\figref{compiling-recovery}),
    and

  \item [recover\_derived] which computes the coordinate of an
    index variable from the variable it derives from and its
    siblings (green arrows in~\figref{compiling-recovery}).

\end{description}

Coordinate recovery may require an expensive search, so we define
an optimization that computes the next coordinate faster than
computing an arbitrary coordinate.  Such tracking code implements
a recurrence through coordinates and requires them to be stored in
order in coordinate hierarchy data structures.  The coordinate
tracking code has two parts, an initialization that finds the
first coordinate and tracking that advances it.  The
initialization is done with the following code generation function
on provenance graphs:

\begin{description}

  \item [recover\_track] which computes the next coordinate (red
    stippled arrow in~\figref{compiling-recovery}).

\end{description}
\figref{generated-code} uses the tracking optimization to track
the \code{i} coordinate.  It starts by finding the first \code{i}
coordinate using a binary search, but then simply advances to the
next \code{i} coordinate when it finds the end of a row segment.

\subsection{Parallel and GPU Code Generation}
\label{sec:parallel-codegen}

Parallelization and vectorization are applied to the high-level
iteration graph IR and their safety can therefore be assured
without heavy analysis.  A parallelization strategy is tagged onto
an index variable and the code generator generates parallel
constructs from it, whether SIMD vector instructions, a parallel
OpenMP loop, or a GPU thread array.  The parallelization command
can easily be extended with other parallelization strategies and
parallel code generators are easy to write as they only
mechanically carry out orders from the higher levels.

It is the responsibility of the parallel code generators to emit
code that safely manage parallel reductions.  We have implemented
two strategies.  The first strategy is to detect data races, by
inspecting whether a reduction is dominated by an index variable
that is summed over, and insert an atomic instruction.  A second
strategy is to separate the loop into worker and a reduction loops
that communicate through a workspace~\cite{kjolstad2019} (e.g., an
array).  The threads in the parallel loop reduce into separate
parts of the workspace.  When they finish, the second loop reduces
across the workspace, either sequentially or in parallel.  We have
implemented this strategy on CPUs using SIMD instructions, and on
GPUs with CUDA warp-level reduction primitives.  It
is also possible to control the reduction strategy at each level
of parallelism to optimize for each level of parallel hardware.
For example, on a GPU we can choose a loop separation strategy
within a warp and atomics across warps.

\section{Scheduling API}
\label{sec:looptransformations}
We expose the sparse transformation primitives as a scheduling API
in TACO, inspired by the Halide system for dense stencil
computations~\cite{ragan-kelley2012}.  The scheduling language is
independent of both the algorithmic language (used to specify
computations) and the format language (used to specify tensor data
structures).  This lets users schedule tensor computations
independently of data structure choice, while ensuring correctness
for the overall algorithm, and further enables efficient execution
on different hardware without changing the algorithm.  We add the
following scheduling APIs to TACO:

\begin{lstlisting}
  IndexStmt reorder(vector<IndexVar> reorderedVars);
  IndexStmt fuse(IndexVar i, IndexVar j, IndexVar f);
  IndexStmt split(IndexVar i, IndexVar i1, IndexVar i2, 
                size_t size);
  IndexStmt divide(IndexVar i, IndexVar i1, IndexVar i2, 
                  size_t size);
  IndexStmt pos(IndexVar i, IndexVar p, Access a);
  IndexStmt coord(IndexVar p, IndexVar i);
  IndexStmt parallelize(IndexVar i, ParallelUnit pu, 
                      OutputRaceStrategy rs);
  IndexStmt unroll(IndexVar i, size_t unrollFactor);
  IndexStmt bound(IndexVar i, BoundType type, 
                  size_t bound);
  IndexStmt precompute(IndexExpr e, IndexVar i, 
                        IndexVar i_pre, Tensor w);
\end{lstlisting}

\begin{figure}[]
  \includegraphics[width=0.9\linewidth]{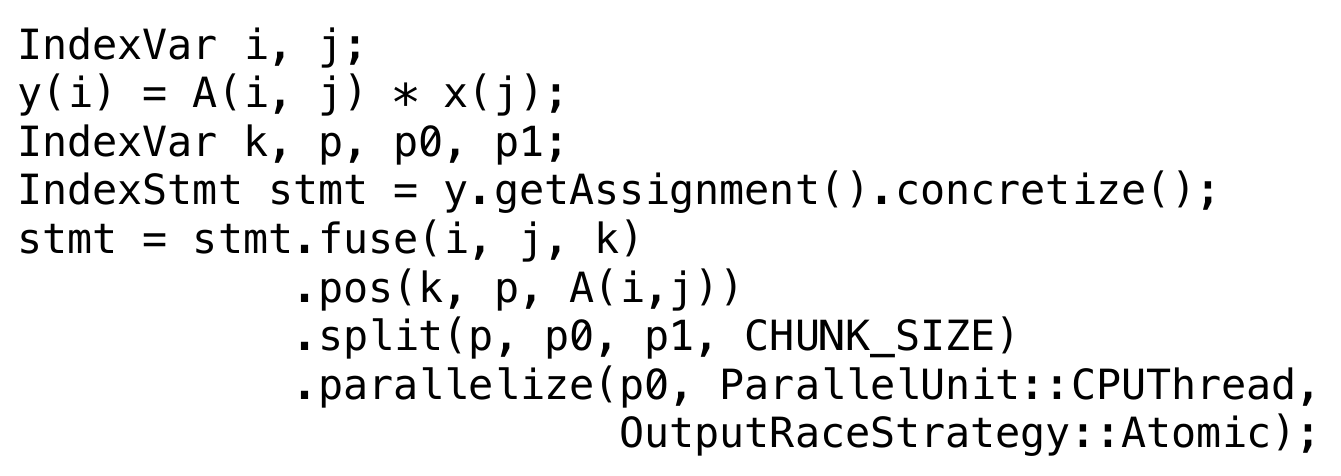}
  \caption {
    The schedule to parallelize SpMV over the nonzeros of the CSR
    matrix is shown. The transformations in this schedule
    correspond to the provenance graph
    in~\figref{provenance-graph} and it generates the code in
    \figref{generated-code}.
    \label{fig:scheduling-example}
  }
\end{figure}

These primitives directly correspond to transformations described
in \secref{transformations}.  The \code{split}, \code{divide}, and
\code{fuse} transformations follow the convention that
derived-from index variables precede newly-derived index variables
in the list of arguments.  \code{reorder} takes a new ordering for
a set of index variables that are directly nested in the iteration order.  
\code{bound} specifies a compile-time constraint on an index variable's
iteration space that allows knowledge of the
size or structured sparsity pattern of the inputs to be 
incorporated during bounds propagation. \code{pos} and \code{coord} create
new index variables in their respective iteration spaces.
\code{pos} requires a tensor access expression as input, that
described the tensor whose coordinate hierarchy to perform a
position cut with respect to.  Specifically, the derived \code{p}
variable will iterate over the tensors' position space at the
level that the \code{i} variable is used in the access expression.
The \code{precompute} transformation is described in prior
work~\cite{kjolstad2019}, but composes with our set of
transformations to allow us to leverage scratchpad memories and
reorder computations to increase locality. The \code{unroll}
primitive unrolls the corresponding loop by a statically-known
integer number of iterations, and finally the \code{parallelize}
primitive is as described in \secref{parallel-codegen}. An example
schedule is shown in Figure \ref{fig:scheduling-example}.  We
first generate an iteration graph by using the \code{concretize}
function and then apply successive transformations using
scheduling language primitives.

\section{Evaluation}
\label{sec:evaluation}
\begin{figure*}
  \center
  \begin{minipage}{0.49\linewidth}
    \includegraphics[width=\linewidth]{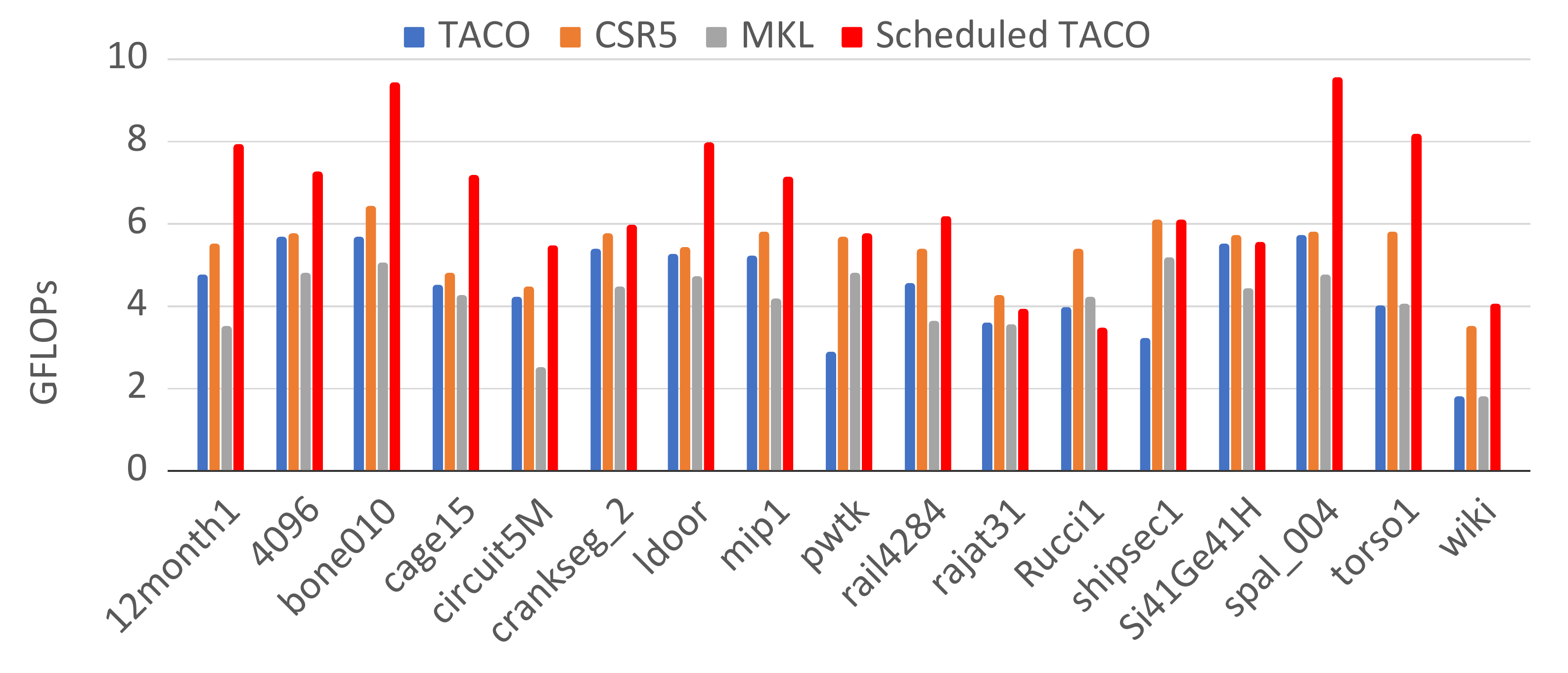}
    \vspace{-7mm}
    \caption {
      \spmv on a CPU.
      \label{fig:results-spmv-cpu}
    }
  \end{minipage}
  \hfill
  \begin{minipage}{0.49\linewidth}
    \includegraphics[width=\linewidth]{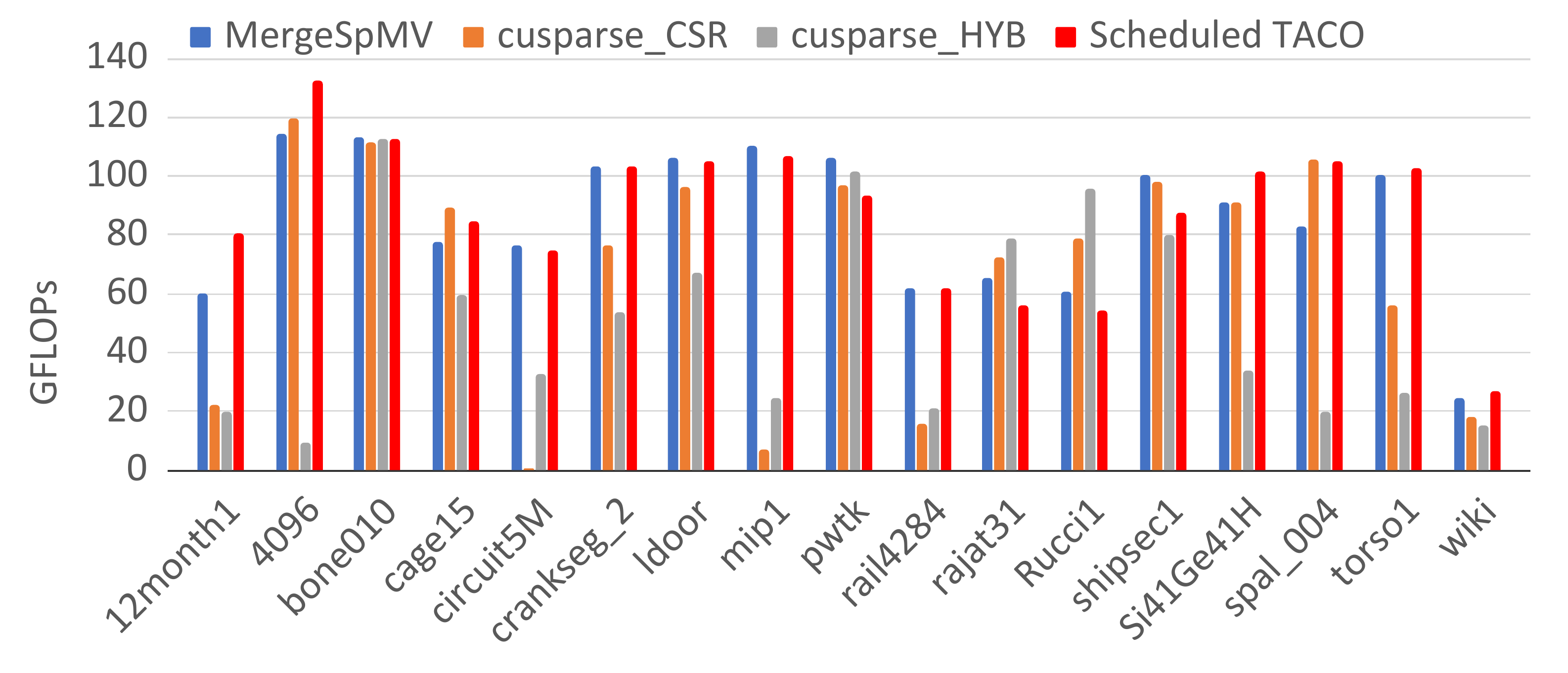}
    \vspace{-7mm}
    \caption {
      \spmv on a GPU.
      \label{fig:results-spmv-gpu}
    }
  \end{minipage}
\end{figure*}

\begin{figure*}
  \center
  \begin{minipage}{0.49\linewidth}
    \includegraphics[width=\linewidth]{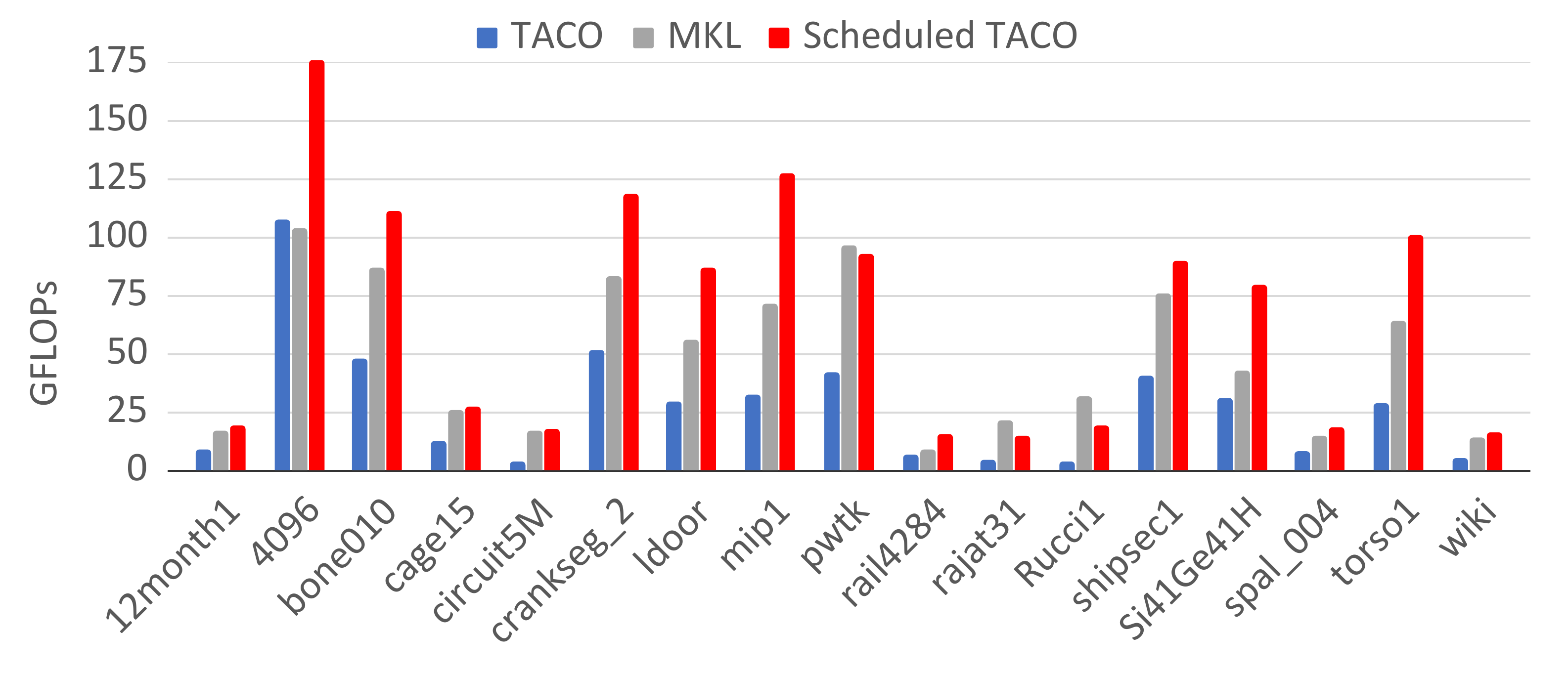}
    \vspace{-7mm}
    \caption {
      SpMM on a CPU.
      \label{fig:results-spmm-cpu}
    }
  \end{minipage}
  \hfill
  \begin{minipage}{0.49\linewidth}
    \includegraphics[width=\linewidth]{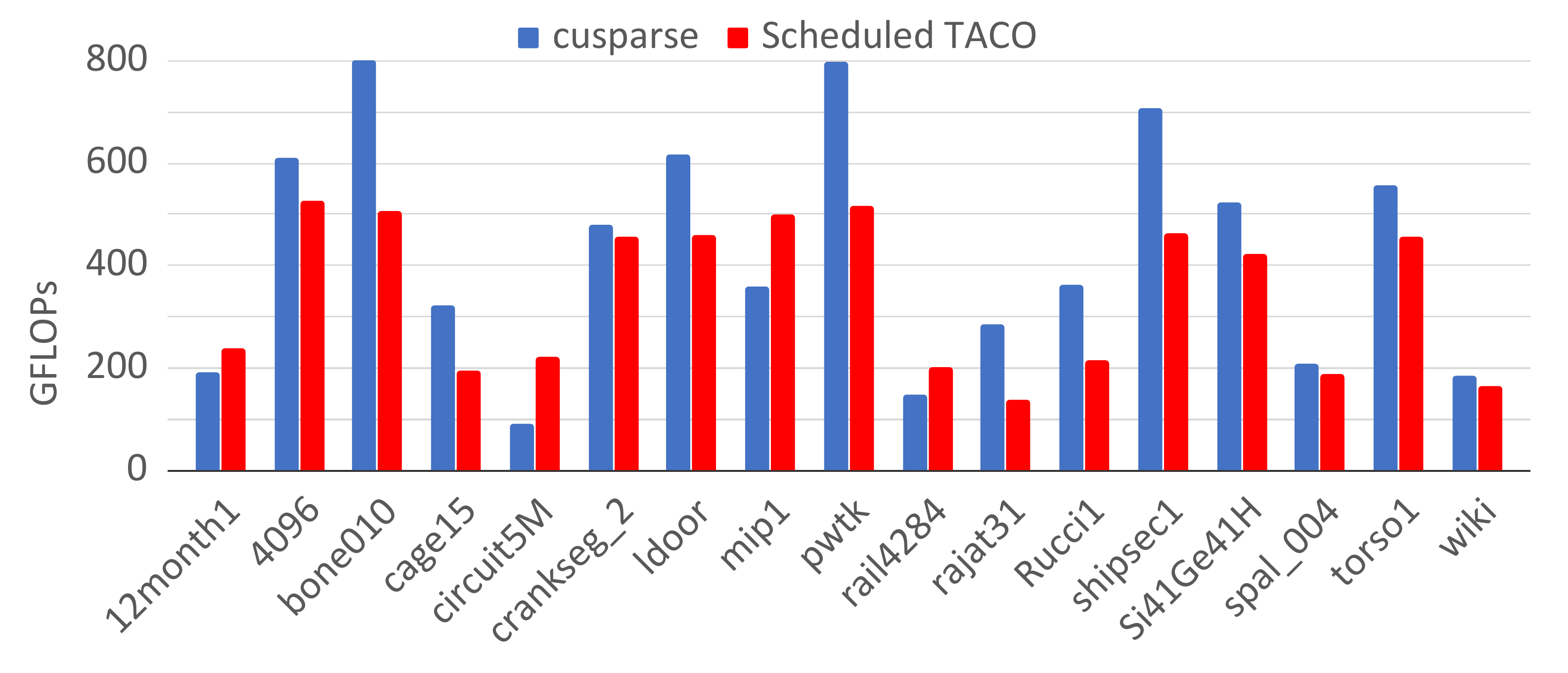}
    \vspace{-7mm}
    \caption {
      SpMM on a GPU.
      \label{fig:results-spmm-gpu}
    }
  \end{minipage}
\end{figure*}

\begin{figure*}
  \center
  \begin{minipage}{0.49\linewidth}
    \includegraphics[width=\linewidth]{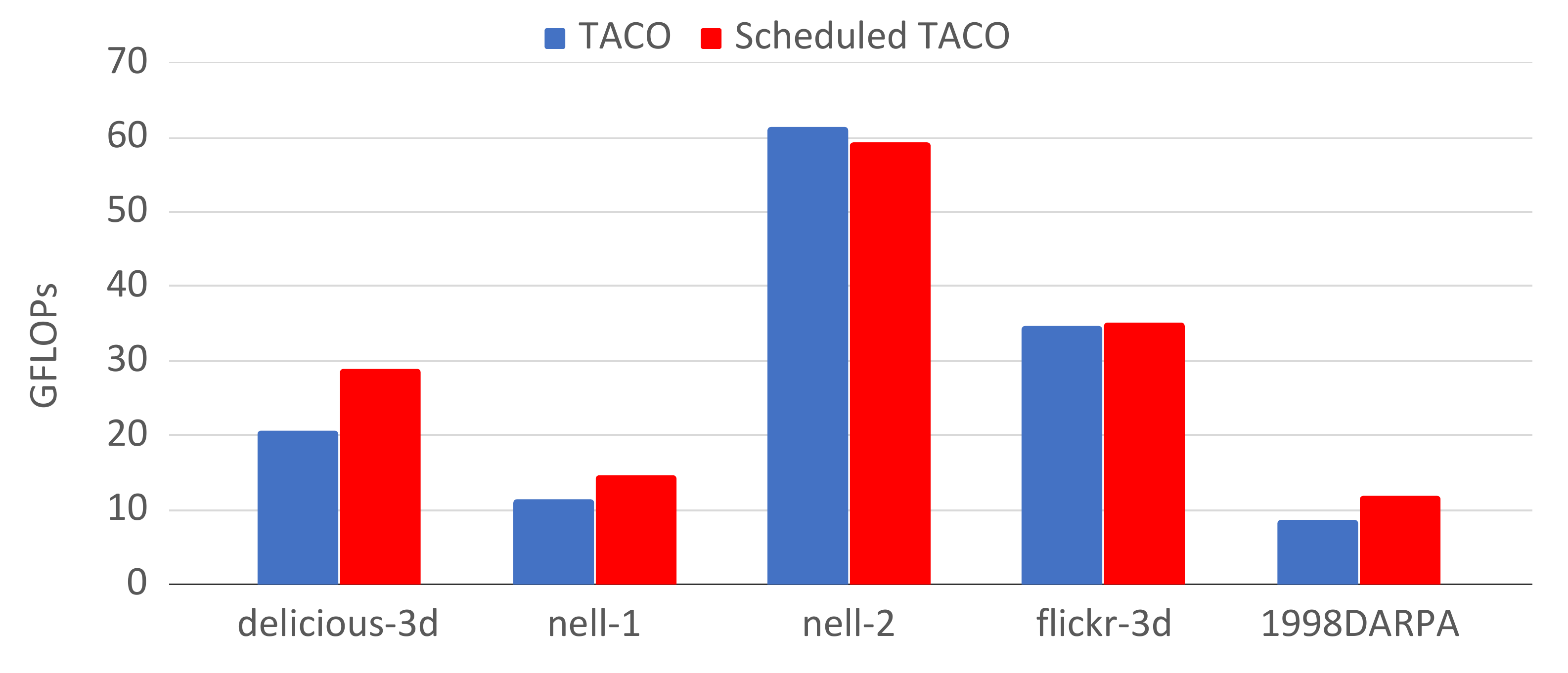}
    \vspace{-7mm}
    \caption {
      MTTKRP on a CPU.
      \label{fig:results-mttkrp-cpu}
    }
  \end{minipage}
  \hfill
  \begin{minipage}{0.49\linewidth}
    \includegraphics[width=\linewidth]{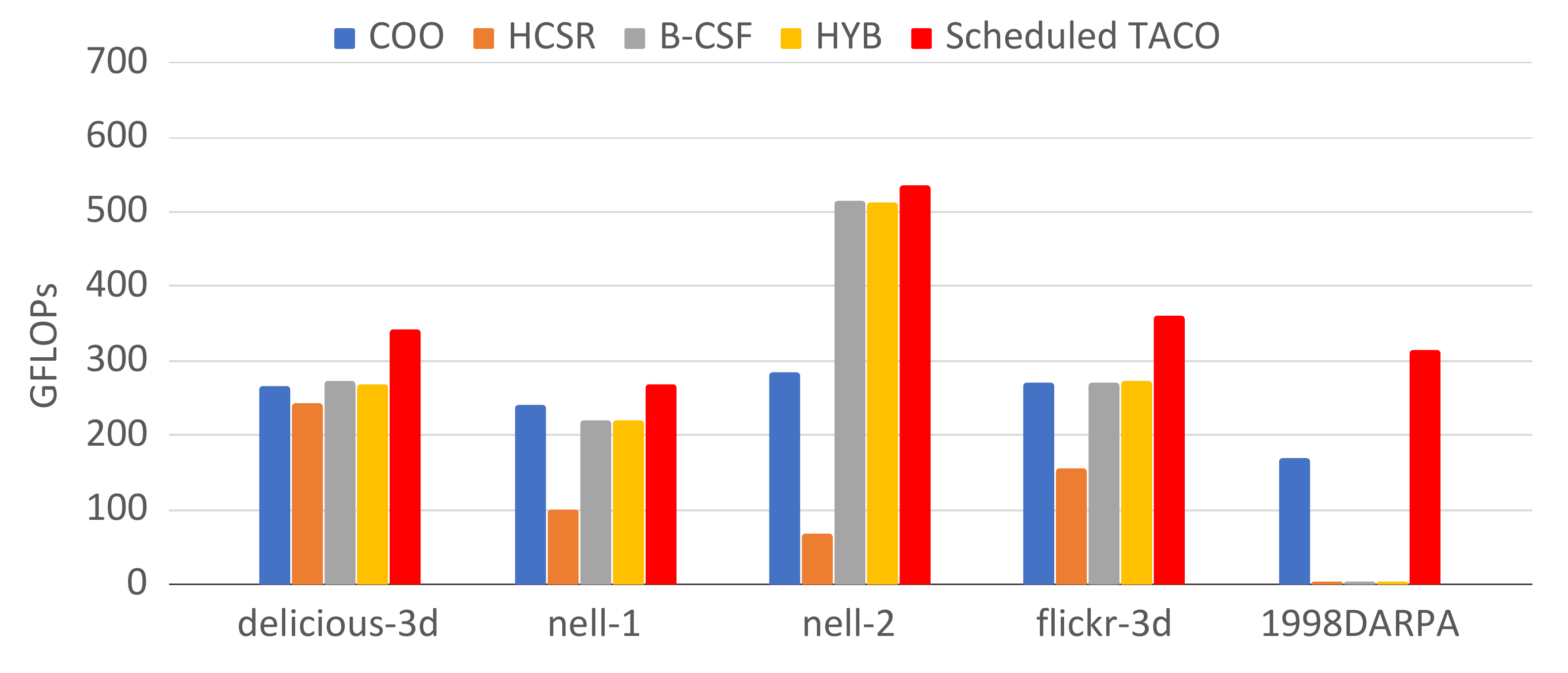}
    \vspace{-7mm}
    \caption {
      MTTKRP on a GPU.
      \label{fig:results-mttkrp-gpu}
    }
  \end{minipage}
\end{figure*}

We carry out experiments to compare the performance of code
generated by our technique with different schedules to
state-of-the-art library implementations of three important
expressions: SpMV, SpMM, and MTTKRP.  We chose these expressions
because they have been heavily studied in the performance
engineering literature.  We stress, however, that the point of our
experiments is not to demonstrate that our transformations produce
kernels with the best possible performance for these specific
expressions.  Rather, the purpose of our transformations is to
apply to any tensor algebra expression.  These three expressions
are therefore stand-ins for any expression, and we seek to
demonstrate that our technique produces code with good performance
that is competitive with hand-optimized kernels.

In addition, we carry out several studies to highlight situations
where the best schedule differs depending on the situation.  For
example, the best CPU and GPU schedules differ, and the best GPU
\spmv schedule depends on whether the computation is load-balanced
or not.  We carry out these studies on the simplest kernels that
are sufficient to make our point.  For most studies this is the
\spmv kernel, except for the locality study
in~\secref{scheduling-for-locality} where we use the SpMM
expression, since it has two dense loops we can tile over.

\subsection{Methodology}

We implement our transformation framework as an extension to the
TACO compiler, which is freely available under the MIT license.
To evaluate it, we compare the performance of code that has been
optimized using our transformations accessed through the
scheduling API to the original TACO system (commit 331188), Intel
MKL~\cite{mkl}, CSR5~\cite{csr5}, cuSPARSE~\cite{cusparse}, and
hand-optimized GPU kernels presented by Nisa et al.~\cite{MTTKRPGPU}.  For the comparative
studies we use the 17 matrices from the SuiteSparse sparse matrix
repository~\cite{suitesparse} used for targeted studies by Merrill
and Garland~\cite{merrill2016} and Steinberger et
al.~\cite{steinberger2017}, and the three tensors from the FROSTT
sparse tensor repository~\cite{frostt} that were used by Kjolstad
et al.~\cite{kjolstad2017}.  We also carry out studies to evaluate
the value of a scheduling language.  Our load-balance study uses
synthetic matrices designed to show at what load imbalance it
makes sense to move to a statically load-balanced kernel.  We have
made all schedules available in the supplementary material.

All CPU experiments are run on a dual-socket, 12-core with 24
threads, 2.5 GHz Intel Xeon E5-2680 v3 machine with 30 MB of L3
cache per socket and 128 GB of main memory, running Ubuntu 18.04.3
LTS.  We compile code that our technique generates using Intel
icpc with \texttt{-O3}, \texttt{-Ofast} and \texttt{-mavx2}.  We
run each experiment with a warm cache 16 times with 8 warm-up runs and report median
serial execution times.  All GPU experiments are run on an NVIDIA
DGX system with 8 V100 GPUs with 32 GB of global memory, 6MB of L2
cache and 128KB of L1 cache per SM (80 SMs), and a bandwidth of
897 GB/s.  We compile the CUDA code that our technique generates
using NVCC v9.0 with \texttt{-O3} and \texttt{--use\_fast\_math}.

\subsection{Comparative Performance}
\label{sec:comparative-performance}

Our first experiments validates that the performance of the code
generated using our transformation framework performs well
compared to other systems, and are shown in
Figures~\ref{fig:results-spmv-cpu}--\ref{fig:results-mttkrp-gpu}.
We evaluate three linear algebra and tensor algebra kernels that
have received a lot of attention from researchers that study
sparse linear and tensor algebra optimization.  These are the
ubiquitous \spmv operation on CSR matrices, the $(\text{sparse
matrix}) \times (\text{dense matrix})$ (SpMM) operation, and the
Matriziced Tensor Times Khatri-Rao Product (MTTKRP) operation.  We
use our scheduling language to optimize each of these for our CPU
and a GPU.  On CPUs we compare the linear algebra kernels to the
TACO Compiler~\cite{kjolstad2017}, the Intel MKL
library~\cite{mkl}, and the CSR5 implementation of
\spmv~\cite{csr5}.  On GPUs we compare the SpMV kernel to
cuSPARSE~\cite{cusparse} using the CSR and hybrid format
implementations of Bell and Garland~\cite{bell2009}, as well as
the Merge-Based SpMV implementation of Merrill and
Garland~\cite{merrill2016}.  For the MTTKRP tensor kernel we
compare to four hand-optimized GPU kernels presented in~\cite{MTTKRPGPU}.
The Scheduled Taco results come from our system. The schedules are
hand-written and included in the supplementary material. Scheduled TACO
uses standard sparse data formats (CSR and CSF) instead of requiring preprocessing.

The results show the kernels generated by our transformations are
competitive with hand-optimized library implementations in most
cases.  One exception is SpMM on GPUs where NVIDIA's cuSPARSE
library performs better.  This library is closed source, however,
so the exact reason is unknown to us.  These results demonstrate
the quality of code generated by our transformations.  Since the
code generator does nothing special for these expressions, it
provides evidence for the performance we may see for other tensor
algebra expressions that have no hand-optimized kernels.

\subsection{Scheduling for GPUs}
\label{sec:scheduling-for-gpus}

Good GPU schedules are different from good CPU schedules, and it
is important to have transformations that let us order operations
to fit the machine at hand.  GPUs are sensitive to the order of
loads and to thread divergence, and typically require more
involved schedules to ensure operations are done in the right
order. For instance, the best parallel CPU \spmv schedule compiled
to and executed on a GPU performs 6.9x worse than the warp-per-row
GPU schedule on a matrix with four million randomly allocated
nonzeros.

The best CPU schedule for the \spmv operation when the matrix is
load-balanced is a simple strip-mining of the outer dense loop to
create parallel blocks, followed by parallelizing the outer loop.
The resulting code assigns a set of rows to each CPU thread
executing in parallel.  The analogous schedule is a disaster on a
GPU.  Since threads in a warp execute separate rows, they cannot
coalesce memory loads.  This results in poor effective memory
bandwidth and thus poor performance for the memory-bound \spmv
kernel.  Furthermore, if there are a different number of nonzeros
on the rows executed by different threads in a warp, then they
will experience thread divergence.

In contrast, more optimized GPU schedules are more carefully
tiled.  The warp-per-row schedule assigns an equal number of
nonzero elements of the row to each thread and uses warp-level
synchronization primitives to efficiently reduce these partial
sums. The optimized SpMV schedule that we use in
\figref{results-spmv-gpu} tiles the position space of the sparse
matrix across threads. We also use a temporary to allow us to
unroll the loop that performs loads and then later use atomic
instructions to store the results in the output. This provides
better memory access patterns and increased instruction-level
parallelism, but makes hand-writing such a kernel difficult. On a
matrix with four million randomly allocated nonzeros, this
increased instruction-level parallelism provided a 36\% speedup
for our optimized schedule over the same schedule without the
temporary or loop unrolling.

\subsection{Scheduling for Load Balance}

This study shows that the best GPU schedules differ for
load-balanced and load-imbalanced computations.  The \spmv
computation demonstrates the issue, as it is sensitive to a skewed
distribution of nonzeros in the matrix.  The challenge, however,
generalizes to any expression with a sparse tensor.

\begin{figure}
  \includegraphics[width=0.9\linewidth]{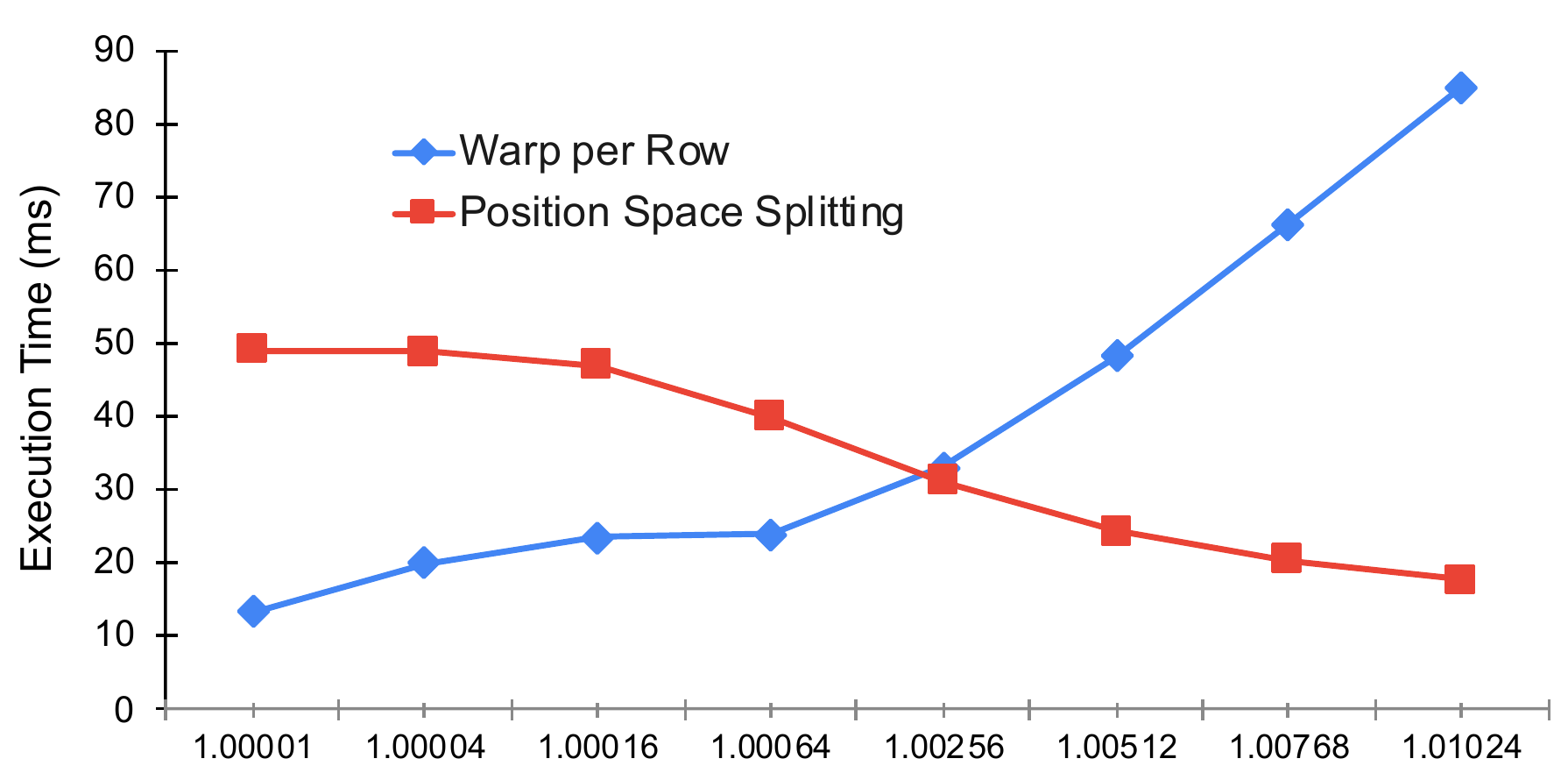}
  \caption {
    Performance comparison of warp-per-row and load-balanced
    \spmv on a matrix with a fixed number of nonzeros as the
    distribution per rows become more skewed.
    \label{fig:load-imbalance}
  }
\end{figure}

The previous section outlined an effective warp-per-row GPU \spmv
schedule where threads in a warp collectively work on a matrix row
at a time.  If distribution of nonzeros across matrix rows is
skewed, this kernel suffers from load imbalance.  The optimized
\spmv schedule used in \figref{results-spmv-gpu} where the two
loops are fused and then split in the position space provides
perfect static load balancing for loads of the sparse tensor at
the cost of overhead from coordinate recovery.
\figref{load-imbalance} shows the performance of the warp-per-row
schedule and the load-balanced position split schedule as the
distribution of nonzeros per row becomes more skewed according to
an exponential function.  The number of nonzeros remains fixed and
rows are randomly shuffled. As expected, the warp-per-thread
schedule performs worse as skew increases, while the load-balanced
schedule benefits from long rows and performs better with
increased skew.  The performance intersects just before the base
of $1.00256$.  For skewed matrices, the load-balanced kernel is
thus preferable.

\subsection{Scheduling for Maximal Parallelism}

Loop fusion to increase the amount of parallelism, despite higher
overhead, can make sense in parallelism-constrained situations.
The warp-per-row GPU \spmv schedule described in
\secref{scheduling-for-gpus} assigns each row to be executed by a
different warp.  For matrices with few rows, however, this may
result in too little parallelism to occupy the GPU.  For such
matrices, fusing before parallelizing the two loops creates
more parallelism.  For example, we executed the \spmv kernels
generated from both schedules on a short and wide $100 \times
100\text{k}$ matrix with $10\text{k}$ nonzeros per row.  As
expected from the experiment design, the warp-per-row kernel had
too little parallelism and the fused kernel ran on average $4.5$
times faster across 10 runs.

\subsection{Scheduling for Locality}
\label{sec:scheduling-for-locality}

Sparse tensor algebra expressions can have dense loops that may be
tiled for better temporal locality.  We demonstrate this for the
SpMM expression.  Since a sparse matrix is multiplied by a sparse
matrix, the resulting kernel has two dense loops that can be
tiled.  We implemented a tiled and an untiled schedule for this
expression and ran the resulting kernels on a $100\text{k} \times
100\text{k}$ sparse matrix with an average of 1000 randomly
distributed nonzeros per row multiplied by a $100\text{k} \times
32$ dense matrix.  Not surprising, the tiled kernel performed
$2$x better.

\section{Related Work}
\label{sec:relatedwork}
There is a long history of optimizations and transformations for
sparse matrices~\cite{vuduc_performance_2002,im_optimizing_2001,
suitesparse} and tensors.  We divide prior work into the
categories described in the following sections.

\paragraph{Sparse compilation}

The work by Bik and Wijshoff~\shortcite{bik1, bik2} was an early
attempt to apply compiler techniques to sparse matrix codes.  They
use a technique called guard encapsulation to turn dense loops
over dense arrays into sparse loops over nonzeros.  The
Bernouilli~\cite{bernoulli} system followed soon thereafter and
lifted matrix codes to relational algebra, that was then optimized
and emitted as sparse code.  Most recent work on compiling and
transforming sparse loops has been done in the context of the
polyhedral model~\cite{strout2012, belaoucha2010,
sparse-polyhedral-framework, pouchet-immutable}.  These generally
employ \textit{inspector-executor} techniques, which combine
run-time inspection of data with compile-time transformations.
Venkat et al.~\cite{chill} use the polyhedral model to turn dense
loops with conditional guards into loops over a sparse matrix,
enabling further optimizations including wavefront
parallelism~\cite{venkat2016automating} and tiling of dense loops
inside sparse codes~\cite{ahmad2016}.  Pugh et al.~\cite{sipr}
designed SIPR, an intermediate representation for sparse
compilation.  Most recently, the TACO compiler generalizes sparse
and dense tensor operations in a variety of formats, automatically
generating code for any computation~\cite{kjolstad2017, chou2018,
kjolstad2019}.  We build on TACO in this work, adding scheduling
and GPU code generation.

\paragraph{Dense tensor compilation} Recent work on dense tensor
algebra has focused on two application areas: quantum chemistry
and machine learning.  While the two areas share some
similarities, different types of tensors and operations are
important in each domain.  The Tensor Contraction
Engine~\cite{auer2006} automatically optimizes dense tensor
contractions, and is developed primarily for chemistry
applications.  Libtensor~\cite{epifanovsky2013} and
CTF~\cite{solomonik2014} are cast tensor contractions as matrix
multiplication by transposing tensors.  In machine learning,
TensorFlow~\cite{tensorflow} and other
frameworks~\cite{pytorch,caffe} combine tensor operations to
efficiently apply gradient descent for learning, and are among the
most popular packages used for deep learning.  TVM~\cite{chen2018}
takes this further by adopting and modifying Halide's scheduling
language to make it possible for machine learning practitioners to
control schedules for dense tensor computations.  Tensor
Comprehensions (TC)~\cite{vasilache2018} is another framework for
defining new deep learning building blocks over tensors, utilizing
the polyhedral model.

\paragraph{Scheduling}

Halide~\cite{halide2012,halide2013} is a widely used domain
specific languages in industrial applications, partially due to
its flexible scheduling language that lets users express how a
high-level algorithm is executed.  Many of our constructs are
inspired by Halide, though we deal with sparse loops while Halide
only considers dense loops.  TVM uses a variant of Halide's
scheduling language, modified for deep learning applications.
Most recently, GraphIt~\cite{graphit} and Taichi~\cite{taichi}
built scheduling languages for graph algorithms and sparse
irregular spatial data structures, respectively.  In the
polyhedral framework, CHiLL~\cite{chill} allows users to specify
sequences of loop transformations, similar to a scheduling
language.

\paragraph{Hand-optimized sparse tensor code}

Finally, several researchers have studied how to manually optimize
sparse linear and tensor algebra code for CPUs and GPUs.  We will
mention a few prominent examples.  The Intel MKL library is a fast
sparse linear algebra library for CPUs that employs some
inspector-executor techniques to choose formats~\cite{mkl}.  Bell
and Garland describe a set of techniques for optimizing \spmv for
several different data structures on GPUs, including the
vectorized kernel we use in our evaluation~\cite{bell2009}.
Furthermore, Merrill and Garland showed how to develop a
load-balanced \spmv implementation by generalizing a parallel
merge algorithm~\cite{merrill2016}.  
The SPLATT library is an
efficient implementation of the MTTKRP kernel~\cite{smith2015b} and
HiCOO explores new coordinate-based formats that further
improves performance~\cite{li2018}.  
Nisa et al. describe techniques for how to effectively parallelize the
MTTKRP kernel for GPUs~\cite{MTTKRPGPU}.
Finally, the Cyclops library
shows how to scale sparse kernels to distributed
machines~\cite{solomonik2017}.

\section{Conclusion}
\label{sec:conclusion}
This paper presented a comprehensive theory of transformations on
sparse iteration spaces.  The resulting transformation machinery
and code generator can recreate tiled, vectorized, parallelized
and load-balanced CPU and GPU codes from the literature, and
generalizes to far more tensor algebra expressions and
optimization combinations.  Furthermore, as the sparse iteration
space transformation machinery works on a high-level intermediate
representation that is independent of target code generators, it
points towards a portable sparse tensor algebra compilation.  With
this work, sparse tensor algebra is finally on the same
optimization and code generation footing as dense tensor algebra
and array codes.

\begin{acks}                            %

We thank Stephen Chou, Michael Pellauer, Ziheng Wang,
Ajay Brahmakshatriya, Albert Sidelnik, Michael Garland,
Rawn Henry, Suzy Mueller, Peter Ahrens, and Yunming Zhang for 
helpful discussion, suggestions, and reviews.
This work was supported by DARPA under Award Number HR0011-18-3-0007; 
the Application Driving Architectures (ADA) Research Center, 
a JUMP Center co-sponsored by SRC and DARPA; 
the U.S. Department of Energy, Office of Science, 
Office of Advanced Scientific Computing Research 
under Award Number DE-SC0018121; the National Science Foundation 
under Grant No. CCF-1533753; and  the Toyota Research Institute. 
Any opinions, findings, and conclusions or recommendations expressed in 
this material are those of the authors and do not necessarily reflect the 
views of the funding agencies.
\end{acks}

\bibliography{bib}

\clearpage
\onecolumn
\appendix
\section{Appendix}
\subsection{SPMV on a CPU (Figure 15)}
\begin{lstlisting}
IndexVar i("i"), j("j");
y(i) = A(i, j) * x(j);

IndexVar i0("i0"), i1("i1"), kpos("kpos"), kpos0("kpos0"), kpos1("kpos1");
IndexStmt stmt = y.getAssignment().concretize();
stmt = stmt.split(i, i0, i1, CHUNK_SIZE)
        .reorder({i0, i1, j})
        .parallelize(i0, ParallelUnit::CPUThread, OutputRaceStrategy::NoRaces);
\end{lstlisting} 

\subsection{SPMV on a GPU (Figure 16)}
\begin{lstlisting}
IndexVar i("i"), j("j");
IndexExpr precomputedExpr = A(i, j) * x(j);
y(i) = precomputedExpr;

IndexVar f("f"), fpos("fpos"), fpos1("fpos1"), fpos2("fpos2");
IndexVar block("block"), warp("warp"), thread("thread");
IndexVar thread_nz("thread_nz"), thread_nz_pre("thread_nz_pre");
TensorVar precomputed("precomputed", 
    Type(Float64, {Dimension(thread_nz)}), taco::dense);
IndexStmt stmt = y.getAssignment().concretize();
stmt = stmt.fuse(i, j, f)
        .pos(f, fpos, A(i, j))
        .split(fpos, block, fpos1, NNZ_PER_TB)
        .split(fpos1, warp, fpos2, NNZ_PER_WARP)
        .split(fpos2, thread, thread_nz, NNZ_PER_THREAD)
        .reorder({block, warp, thread, thread_nz})
        .precompute(precomputedExpr, thread_nz, thread_nz_pre, precomputed)
        .unroll(thread_nz_pre, NNZ_PER_THREAD)
        .parallelize(block, ParallelUnit::GPUBlock, OutputRaceStrategy::IgnoreRaces)
        .parallelize(warp, ParallelUnit::GPUWarp, OutputRaceStrategy::IgnoreRaces)
        .parallelize(thread, ParallelUnit::GPUThread, OutputRaceStrategy::Atomics);
\end{lstlisting} 

\subsection{SPMM on a CPU (Figure 17)}
\begin{lstlisting}
IndexVar i("i"), j("j"), k("k");
C(i, k) = A(i, j) * B(j, k);

IndexVar i0("i0"), i1("i1"), kbounded("kbounded"), k0("k0"), k1("k1");
IndexVar jpos("jpos"), jpos0("jpos0"), jpos1("jpos1");
IndexStmt stmt = C.getAssignment().concretize();
stmt = stmt.split(i, i0, i1, CHUNK_SIZE)
        .pos(j, jpos, A(i,j))
        .split(jpos, jpos0, jpos1, UNROLL_FACTOR)
        .reorder({i0, i1, jpos0, k, jpos1})
        .parallelize(i0, ParallelUnit::CPUThread, OutputRaceStrategy::NoRaces)
        .parallelize(k, ParallelUnit::CPUVector, OutputRaceStrategy::IgnoreRaces);
\end{lstlisting} 

\subsection{SPMM on a GPU (Figure 18)}
\begin{lstlisting}
IndexVar i("i"), j("j"), k("k");
C(i, k) = A(i, j) * B(j, k);

IndexVar f("f"), fpos("fpos"), block("block"), fpos1("fpos1"), warp("warp");
IndexVar nnz_pre("nnz_pre"), nnz("nnz");
IndexVar dense_val_unbounded("dense_val_unbounded");
IndexVar dense_val("dense_val"), thread("thread"), thread_nz("thread_nz");
TensorVar precomputed("precomputed", 
    Type(Float64, {Dimension(nnz)}), taco::dense);
IndexStmt stmt = C.getAssignment().concretize();
stmt = stmt.reorder({i, j, k})
        .fuse(i, j, f)
        .pos(f, fpos, A(i, j))
        .split(fpos, block, fpos1, NNZ_PER_TB)
        .split(fpos1, warp, nnz, NNZ_PER_WARP)
        .split(k, dense_val_unbounded, thread, WARP_SIZE)
        .bound(dense_val_unbounded, dense_val, 1, BoundType::MaxExact)
        .reorder({block, warp, dense_val, thread, nnz})
        .parallelize(block, ParallelUnit::GPUBlock, OutputRaceStrategy::IgnoreRaces)
        .parallelize(warp, ParallelUnit::GPUWarp, OutputRaceStrategy::IgnoreRaces)
        .parallelize(thread, ParallelUnit::GPUThread, OutputRaceStrategy::Atomics);
\end{lstlisting} 

\subsection{MTTKRP on a CPU (Figure 19)}
\begin{lstlisting}
IndexVar i("i"), j("j"), k("k");
A(i,j) = B(i,k,l) * C(k,j) * D(l,j);

IndexVar ipos("ipos"), ipos0("ipos0"), ipos1("ipos1");
IndexStmt stmt = A.getAssignment().concretize();
stmt = stmt.pos(i, ipos, B(i, k, l))
      .split(ipos, ipos0, ipos1, CHUNK_SIZE)
      .reorder({ipos0, ipos1, k, l, j})
      .parallelize(ipos0, ParallelUnit::CPUThread, OutputRaceStrategy::NoRaces);
\end{lstlisting} 

\subsection{MTTKRP on a GPU (Figure 20)}
\begin{lstlisting}
IndexVar i("i"), j("j"), k("k");
A(i,j) = B(i,k,l) * C(k,j) * D(l,j);

IndexVar kl("kl"), f("f"), fpos("fpos"), block("block"), fpos1("fpos1");
IndexVar warp("warp"), nnz("nnz"), dense_val("dense_val");
IndexVar dense_val_unbounded("dense_val_unbounded"), thread("thread");
IndexStmt stmt = A.getAssignment().concretize();
stmt = stmt.reorder({i,k,l,j})
      .fuse(k, l, kl)
      .fuse(i, kl, f)
      .pos(f, fpos, B(i, k, l))
      .split(fpos, block, fpos1, NNZ_PER_TB)
      .split(fpos1, warp, nnz, NNZ_PER_WARP)
      .split(j, dense_val_unbounded, thread, WARP_SIZE)
      .bound(dense_val_unbounded, dense_val, 1, BoundType::MaxExact)
      .reorder({block, warp, dense_val, thread, nnz})
      .parallelize(block, ParallelUnit::GPUBlock, OutputRaceStrategy::IgnoreRaces)
      .parallelize(warp, ParallelUnit::GPUWarp, OutputRaceStrategy::IgnoreRaces)
      .parallelize(thread, ParallelUnit::GPUThread, OutputRaceStrategy::Atomics);
\end{lstlisting} 

\subsection{SpMV Thread per Row on GPU (Section 8.3)}
\begin{lstlisting}
IndexVar i("i"), j("j");
y(i) = A(i, j) * x(j);

IndexVar block("block"), warp("warp"), thread("thread");
IndexVar thread_nz("thread_nz"), i1("i1"), jpos("jpos");
IndexVar block_row("block_row"), warp_row("warp_row");
IndexStmt stmt = y.getAssignment().concretize();
stmt = stmt.split(i, block, thread, ROWS_PER_TB)
      .parallelize(block, ParallelUnit::GPUBlock, OutputRaceStrategy::NoRaces)
      .parallelize(thread, ParallelUnit::GPUThread, OutputRaceStrategy::NoRaces);
\end{lstlisting}

\subsection{SpMV Warp per Row on GPU (Section 8.3)}
\begin{lstlisting}
IndexVar i("i"), j("j");
IndexExpr precomputedExpr = A(i, j) * x(j);
y(i) = precomputedExpr;

IndexVar block("block"), warp("warp"), thread("thread"), thread_nz("thread_nz");
IndexVar i1("i1"), jpos("jpos"), block_row("block_row"), warp_row("warp_row");
TensorVar precomputed("precomputed", 
    Type(Float64, {Dimension(thread_nz)}), taco::dense);
IndexStmt stmt = y.getAssignment().concretize();
stmt = stmt.split(i, block, block_row, ROWS_PER_TB)
        .split(block_row, warp_row, warp, BLOCK_SIZE / WARP_SIZE)
        .pos(j, jpos, A(i, j))
        .split(jpos, thread_nz, thread, WARP_SIZE)
        .reorder({block, warp, warp_row, thread, thread_nz})
        .parallelize(block, ParallelUnit::GPUBlock, OutputRaceStrategy::IgnoreRaces)
        .parallelize(warp, ParallelUnit::GPUWarp, OutputRaceStrategy::IgnoreRaces)
        .parallelize(thread, ParallelUnit::GPUThread, OutputRaceStrategy::Temporary);
\end{lstlisting} 

\subsection{SpMV on GPU with no Unrolling (Section 8.3)}
\begin{lstlisting}
IndexVar i("i"), j("j");
IndexExpr precomputedExpr = A(i, j) * x(j);
y(i) = precomputedExpr;

IndexVar f("f"), fpos("fpos"), fpos1("fpos1"), fpos2("fpos2");
IndexVar block("block"), warp("warp"), thread("thread"), thread_nz("thread_nz");
IndexStmt stmt = y.getAssignment().concretize();
stmt = stmt.fuse(i, j, f)
        .pos(f, fpos, A(i, j))
        .split(fpos, block, fpos1, NNZ_PER_TB)
        .split(fpos1, warp, fpos2, NNZ_PER_WARP)
        .split(fpos2, thread, thread_nz, NNZ_PER_THREAD)
        .reorder({block, warp, thread, thread_nz})
        .parallelize(block, ParallelUnit::GPUBlock, OutputRaceStrategy::IgnoreRaces)
        .parallelize(warp, ParallelUnit::GPUWarp, OutputRaceStrategy::IgnoreRaces)
        .parallelize(thread, ParallelUnit::GPUThread, OutputRaceStrategy::Atomics);
\end{lstlisting}

\subsection{SpMM on CPU Tiled (Section 8.6)}
\begin{lstlisting}
IndexVar i("i"), j("j"), k("k");
C(i, k) = A(i, j) * B(j, k);

IndexVar i0("i0"), i1("i1");
IndexVar jpos("jpos"), jpos0("jpos0"), jpos1("jpos1");
IndexStmt stmt = C.getAssignment().concretize();
stmt = stmt.pos(j, jpos, A(i,j))
      .split(jpos, jpos0, jpos1, UNROLL_FACTOR)
      .reorder({i, jpos0, k, jpos1});
\end{lstlisting} 

\subsection{SpMM on CPU No Tiling (Section 8.6)}
\begin{lstlisting}
IndexVar i("i"), j("j"), k("k");
C(i, k) = A(i, j) * B(j, k);

IndexStmt stmt = C.getAssignment().concretize();
\end{lstlisting}

\end{document}